\documentclass[]{svjour3}

\makeatletter
\let\cl@chapter\undefined
\makeatletter

\usepackage[applemac]{inputenc}
\usepackage[T1]{fontenc}
\usepackage{amsmath}
\usepackage{bm}
\usepackage{amsfonts}
\usepackage{amssymb,amsbsy,mathrsfs}
\usepackage[usenames,dvipsnames,svgnames,table]{xcolor}
\usepackage{nicefrac}
\usepackage{graphicx}
\usepackage{enumerate}
\usepackage{commath}
\usepackage{ulem}
\usepackage{cleveref}
\usepackage[numbers]{natbib}
\usepackage{soul}
\usepackage{mymacros_10Sep19}
\usepackage{comment}
\usepackage[outdir=./]{epstopdf}

\usepackage{fullpage,lmodern}

\usepackage[inline,final]{showlabels,rotating}

\crefalias{section}{appendix}
\crefalias{Section}{Appendix}

\usepackage{authblk}
\usepackage{booktabs}

\newcommand{\pa}{\partial}
\newcommand{\xb}{\bm{x}}

\newcommand{\nb}{\bm{n}}

\newcommand{\be}[1]{
\begin{equation}
\expandafter\label{eq:#1}
}
\newcommand{\ee}{\end{equation}}
\newcommand{\eq}[1]{Eq.~\eqref{eq:#1}}
\newcommand{\eqs}[1]{Eqs.~\eqref{eq:#1}}

\newcommand{\bfig}[1]{
\begin{figure}
\expandafter\label{fig:#1}
}
\newcommand{\efig}{\end{equation}}
\newcommand{\fig}[1]{Fig.~\ref{fig:#1}}

\newcommand{\Die}[1]{$10^{{#1}}$}

\newcommand{\grad}{\nabla}
\newcommand{\mV}[1]{\left\langle #1 \right\rangle}
\newcommand{\vm}{\bm{v}}
\newcommand{\bo}{\beta}
\newcommand{\ao}{\alpha}
\newcommand{\ue}{\bm{u}}

\newcommand{\por}{\Omega}
\newcommand{\pore}{\por\times (0,\infty)^2}
\newcommand{\porb}{\Gamma}
\newcommand{\poreb}{\porb\times(0,\infty)^2}

\newcommand{\tpore}{\tilde{\por}\times (0,\infty)^2}

\newcommand{\tporeb}{\tilde{\porb}\times(0,\infty)^2}

\newcommand{\fe}{f}
\newcommand{\Da}[1]{\mathrm{Da}_\mathrm{#1}}
\newcommand{\diff}{ D }
\newcommand{\tdiff}{\widetilde{\diff}}
\newcommand{\diffref}{D_0}
\newcommand{\tfe}{\tilde{\fe}}
\newcommand{\tvm}{\tilde{\vm}}
\newcommand{\psize}{r}
\newcommand{\tpsize}{\tilde{\psize}}
\newcommand{\refpsize}{\ell}
\newcommand{\cfbrown}{K_{\mathrm{B}}}
\newcommand{\brown}{\alpha_{\mathrm{B}}}
\newcommand{\tbrown}{\tilde{\alpha}_{\mathrm{B}}}
\newcommand{\kshear}{\alpha_{\mathrm{S}}}
\newcommand{\cfkshear}{K_{\mathrm{S}}}
\newcommand{\tkshear}{\tilde{\alpha}_{\mathrm{S}}}
\newcommand{\shear}{\sigma}
\newcommand{\tshear}{\tilde{\sigma}}
\newcommand{\peclet}{P\'{e}clet\ }
\newcommand{\Dam}{Damk{\"o}hler\ }
\newcommand{\paccl}{\mathrm{\bfa}}
\newcommand{\tpaccl}{\tilde{\mathrm{\bfa}}}
\newcommand{\diffsett}{\alpha_{\mathrm{D}}}
\newcommand{\tdiffsett}{\tilde{\alpha}_{\mathrm{D}}}
\newcommand{\cfdiffsett}{K_{\mathrm{D}}}
\newcommand{\uflow}{\bfu}
\newcommand{\tuflow}{\tilde{\bfu}}

\newcommand{\taggr}{\widetilde{A}}
\newcommand{\aggr}{A}
\newcommand{\tdeath}{\widetilde{B}}
\newcommand{\death}{B}
\newcommand{\tsurf}{\widetilde{C}}
\newcommand{\surf}{C}

\newcommand{\ndimbr}{\Theta}

\newcommand{\tDdistr}{\tilde{\mathcal{A}}}
\newcommand{\GDdistr}{\mathfrak{A}}
\newcommand{\tGDdistr}{\tilde{\mathfrak{A}}}
\newcommand{\mG}[1]{\aver{ #1  }^\Gamma}

\newcommand{\pe}{\mathrm{Pe}}
\newcommand{\da}{\mathrm{Da}}
\renewcommand{\st}{\mathrm{St}}
\newcommand{\fr}{\mathrm{Fr}}

\begin{document}
\graphicspath{{./}{./figures/}{./Figures/}}

\DeclareGraphicsExtensions{.pdf,.png,.jpg}

\title{Population balance models for particulate flows in porous media: breakage and shear-induced events}

\author{Matteo Icardi \and
    Nicodemo Di Pasquale \and
    Eleonora Crevacore \and
    Daniele Marchisio \and
    Matthaus U. Babler
}

\institute{M. Icardi \at
        School of Mathematical Sciences, University of Nottingham, UK
           \and
           N. Di Pasquale \at
              Department of Chemical Engineering and Analytical Science, University of Manchester, UK
          \and
          E. Crevacore \at 
          Dipartimento di Scienze Matematiche ``G.L. Lagrange'', Politecnico di Torino, Italy
          \and
          D. Marchisio \at
          DISAT, Politecnico di Torino, Italy
          \and
          M.U. Babler \at
          Dept. Chemical Engineering, KTH Royal Insitute of Technology, SE-10044 Stockholm, Sweden
}

\maketitle

\begin{abstract}
 Transport and particulate processes are ubiquitous in environmental, industrial and biological applications, often involving complex geometries and porous media. In this work we present a general population balance model for particle transport at the pore-scale, including aggregation, breakage and surface deposition. The various terms in the equations are analysed with a dimensional analysis, including a novel collision-induced breakage mechanism, and split into one- and two-particles processes. While the first are linear processes, they might both depend on local flow properties (e.g. shear). This means that the upscaling (via volume averaging and homogenisation) to a macroscopic (Darcy-scale) description requires closures assumptions.
 We discuss this problem and derive an effective macroscopic term for the shear-induced events, such as breakage caused by shear forces on the transported particles. We focus on breakage events as prototype for linear shear-induced events and derive upscaled breakage frequencies in periodic geometries, starting from non-linear power-law dependence on the local fluid shear rate. Results are presented for a two-dimensional channel flow and a three dimensional regular arrangement of spheres, for arbitrarily fast (mixing-limited) events. Implications for linearised shear-induced collisions are also discussed.
 This work lays the foundations of a new general framework for multiscale modelling of particulate flows.
 \end{abstract}


\section{Introduction}

The evolution of a population of particles can be described using the Population Balance Equation (PBE) \citep{Ramkrishna2000}, which solves for the distribution of particle sizes in presence of phenomena such as breakup and aggregation. While the PBE framework was successfully applied in different chemical processes, such as the precipitation of polymer nanoparticles \citep{DiPasquale2012,Lavino2017,Lattuada2016}, aggregation of colloids \citep{sadegh2018mechanisms,Serra1997,Serra1998}, or the description of foams and bubble evolution \citep{Zitha2010,Buffo2012}, limited works studying particulate processes in porous media, via the PBE, exist. While the PBE itself can be thought as a meso-scale description where the particles are described in a statistical ensemble conditioned on their size, it has wide applicability, thanks to its possible ramifications to different ``scales''. 
For example, we can describe the evolution of complex systems (e.g., polymer macromolecules) if we couple the PBE framework with microscopic modelling, such as Molecular Dynamics simulations \citep{DiPasquale2013,DiPasquale2014pcl} or discrete element methods \citep{frungieri2020shear}; or we can couple its constituent terms (e.g. brekage and collision frequencies) with local flow properties obtained from Computational Fluid Dynamics (CFD) calculations \citep{DiPasquale2012,Lavino2017}, obtaining a very precise description of the system considered. 

In the presence of complex geometries and porous materials, the detailed resolution of the flow structures, while in principle possible \citep{Marcato2021}, it is often impractical and upscaled porous media models need to be introduced. While the derivation and calibration of single- and multi-phase flow and transport models have been widely studied in the porous media and fluid dynamics community \cite{Boccardo2017,Icardi2020}, there is currently no general population balance theory at the Darcy/continuum scale. \citet{Krehel2015} studied the upscaling of a discrete number of particle sizes and derived upscaled equations in the assumption of slow aggregation kinetics. Several other authors \citep{won2021upscaling,Boccardo2019,seetha2017upscaling} have studied the effect of poly-dispersity on transport and retention in porous media without including explicitly the particulate processes. \citet{bedrikovetsky2008upscaling} and subsequent works from the same authors developed micro- and macro-models based on discrete PBE concepts to explain various porous media phenomena but with no direct link to the pore-scale solution. Population balances have been applied at the Darcy-scale \citep{nassar2015development,Patzek1988} in a phenomenological way and parameters estimated experimentally.

The objective of this paper is, on one hand, to provide a thorough description of aggregation, breakage, and other particulate processes described by the PBE, while, on the other hand, laying the foundations of their multiscale analysis in heterogeneous and porous media.  While the development of an effective macroscopic population balance model is the ultimate step, the starting point is the understanding of the fundamental processes and the emergent upscaled mechanisms. This is achieved by analysing the {collision events} and their dependence on particle and flow properties. To this aim, we  provide, for the first time, a general dimensional analysis of all terms, including one- and two-particle processes, and we propose an upscaling methodology to extract macroscopic effective frequency kernels with a rigorous bottom-up approach.

Regarding the upscaling, in this work we limit ourselves to shear-induced breakage events in periodic media. 
A similar approach could also be applied for competitive events, considering multiple collision or breakage mechanisms. Similarly to the usual approach in population balance studies, the overall reaction rate (or collision kernel) can be approximated as the sum of the contributions of the separate mechanisms. This is however not necessarily true for non-linear (i.e. two-particle) events and when scales are not fully separable. Extensions of the upscaling methodology presented here, including non-linear aggregation terms and more complex geometrical and physical models, are currently under development and will be subject of future investigations.

This paper is organised as follows. We first present a discussion of the PBE with particular emphasis on the aggregation and breakage mechanism, presented here in terms of single- and two-particles events. We will give a detailed derivation of the dimensional analysis of the PBE. We then move to describe the motion of a suspension of poly-dispersed particles through a porous medium filled with an incompressible fluid undergoing creeping-flow conditions, where we couple the non-dimensional PBE with the spatial averages over the pore domain. We then specialise the discussion to the breakage of the particle showing the relevant equations and derivations which include the collision models and the upscaling step. We then apply the framework we derived to two test cases, a two-dimensional channel flow and a three-dimensional face-centred cubic domain. We then presents some results and draw some conclusions.

\section{Particulate flow model}

Let us consider the motion of a suspension of a poly-dispersed population of particles through a porous medium filled with an incompressible fluid undergoing creeping-flow conditions. Let $\por$ be the fluid domain in a Representative Elementary Volume (REV) of a periodic porous medium, whose internal boundary is denoted as $\Gamma$.

While the flow through porous media can be described by the Darcy's equation, we start from the microscopic fluid dynamics description of an incompressible fluid in creeping flow conditions. Therefore, we introduce the dimensionless steady Stokes flow  (with viscous scaling)  \citep{auriault1995taylor}:
\begin{align}\label{eq:campoDiMoto}
&\dfrac{\mu U}{P L}\Delta \ue = \nabla p &  \mathrm{in }\quad \por, \nonumber \\
&\nabla \cdot \ue = 0  &  \mathrm{in} \quad \por, \\
&\ue = \bm{0} &   \mathrm{on} \quad \porb, \nonumber
\end{align}
where $\ue$ is the (dimensionless) fluid velocity and $p$ is the (dimensionless) pressure; $\mu$ is the viscosity, $P,U,L$ are reference pressure, velocity and length, respectively.
We consider a solid matrix composed of a collection of identical grains and choose the characteristic length to be  equal to the grains diameter. 
No-slip conditions ($\ue=0$) are imposed on the grains surface $\Gamma$, while the other boundary conditions are not specified in \eq{campoDiMoto} as they depend on the specific scenario.

Now, we consider the transport of a population of particles, described by the (dimensional) Particle Size Distribution (PSD) $\tfe(\tilde{t},\tilde{\bfx},\tpsize)$, 
in the pore space, that undergo aggregation, breakage and surface deposition. 
$\tfe(\tilde{t},\tilde{\bfx},\tpsize)$ represents the number density function of particles with dimensional size $\tpsize$, at the time $\tilde{t}$ in the position $\tilde{\bfx}$. In order to avoid the definition of too many symbols, as a convention in this work, a symbol under the tilde ($\tilde{\cdot}$) represents a dimensional quantity. The same symbol without the tilde represents the non-dimensional version of the same quantity. We start by defining the following dimensionless quantities: a reference particle size $\refpsize$, a reference diffusion coefficients of the particles $\diffref$, and a reference distribution $F$. We note that, if $\tfe$ represents the numerical distribution of the size of the particle per unit volume it has dimension of the inverse of a length to the power of four, i.e. $\sqp{\nicefrac{1}{L^3 \refpsize}}$.  We note here that our problem therefore includes the definition of two different reference lengths, a ``meso-scopic'' one, referring to the pore-scale REV and a ``micro-scopic'' one referring to the particles. A third ``macro-scopic`` scale arises when averaging over the REV onto a larger continuum pore domain. In this work we focus on the micro- and meso-scales while we will obtain averaged quantities that can then be used at the macro-scale.

The evolution of $\tfe(\tilde{t},\tilde{\bfx},\tpsize)$ is described by the Population Balance Equation (PBE), which reads as \citep{Marchisio2013,Ramkrishna2000}:
\begin{subequations}
\label{eq:dimpbe}
\begin{align}
&\pa_{\tilde{t}}  \tfe +  \tilde{\nabla} \cdot \left(\tilde{\vm}  \tilde{f} \right) -  \tdiff  \widetilde{\Delta}  \tfe =
\taggr \left( \tfe,  \tfe,\tpsize \right) +   \tdeath_1 \cip{ \tfe, \tpsize } + \tdeath_2 \cip{ \tfe,\tfe, \tpsize } & &\mathrm{in }\, \tpore, \label{eq:pbeB} \\
&\nb\cdot\del{\, \tvm\tfe - \tdiff \widetilde{ \grad}  \tfe} = \tsurf(\tfe, \tpsize) & &\mathrm{on }\, \tporeb, \label{eq:pbeG}
\end{align}
\end{subequations}
where we dropped all the dependencies of $\tfe$ for the sake of readability. 
The terms on the right hand side of Eq. (\ref{eq:dimpbe}) describe the particulate processes (aggregation, breakup, deposition) where the double appearance of $\tfe$ in these terms indicates two-particle processes while, accordingly, the single appearance of $\tfe$ indicates one-particle processes.
Before describing in detail the different terms appearing in \cref{eq:dimpbe} we rewrite it in dimensionless form, dividing \cref{eq:pbeB} by $ \frac{F \diffref}{L^2} $ and \cref{eq:pbeG} by $ \frac{F \diffref}{L} $.
\begin{align}\label{eq:adimpbe}
& \pard{f}{t}  +  \pe  \nabla \cdot \cip{\fe\vm}  -   \diff \Delta \fe =
A  +  B_1 + B_2  & &\mathrm{in }\, \pore, \nonumber \\
&\nb\cdot\del{\pe\, \vm\fe -  \diff \grad \fe} =  C & &\mathrm{on }\, \poreb,
\end{align}
where we have used the non-dimensionalisation variables shown in \cref{tab:nondim}. 
It is important to notice here that both  the diffusion $\tdiff$ and $\tilde\vm$ may  depend on particle size and therefore, after dividing by a reference diffusion coefficients and velocity (related to the reference particle size), both dimensionless functions $\diff$ and $\vm$ remain in the equation.

{
\renewcommand\arraystretch{1.5}
\begin{table}[htpb]
\begin{center}
\begin{tabular}{p{2.5cm}| p{2.5cm} | p{2.5cm} }
	$\tfe = F \fe$ &  $\tilde{\bfx} = L \bfx $ & $ \tilde{t} = L^2\diffref^{-1} t$ \\
	$\tvm =  U \vm $  & $\tdiff    =  \diffref \diff$  &   $\tpsize = \refpsize \psize$ \\
       $\widetilde{\nabla}   =L^{-1}  \nabla$  & $\widetilde{\Delta} =   L^{-2}  \Delta$ & $\pard{}{\tilde{t}}  =  \diffref L^{-2}  \pard{}{t}$ \\
        $\taggr = F \diffref L^{-2} \aggr$ & $\tdeath_1  = F \diffref L^{-2} \death_1$ & $\tdeath_2  = F \diffref L^{-2} \death_2$ \\     $\tsurf  = F \diffref L^{-1} \surf$     & $\pe = UL\diffref^{-1}$ & \\
\end{tabular}
\caption{Definition of non-dimensional quantities used in the derivation of the PBE.}
\label{tab:nondim}
\end{center}
\end{table}
}

To obtain \cref{eq:adimpbe,eq:dimpbe}, we have made the following modelling assumptions.

\paragraph{Particle velocity:}
$\vm$ is the dimensionless velocity of particles of size $\tpsize$. We assume this to be a given function of both the velocity field $\uflow$ and the particle size, i.e., $\vm=\ue+\ue_{\psize,\mathrm{rel}}$, where $\ue_{\psize,\mathrm{rel}}$ is the slip velocity depending on the particle size.
For small Stokes number we can use  the approximated expression reported in \citep{Ferry2003} for the relative velocity $\tilde{\ue}_{\psize,\mathrm{rel}}$:
\begin{equation}\label{eq:velpart}
    \tvm=\tuflow+\tau(\tpsize)\tilde{\paccl} 
\end{equation}
which can be rewritten in non-dimensional form as:
\begin{align}
 \vm = \frac{\tvm}{U} =\frac{\tuflow}{U} + \frac{\tau(\tpsize)}{U}\tpaccl = \uflow + \dfrac{\st}{\fr^2}\, \paccl\,\psize^2\,,
\end{align}
where $\tau(\tpsize)=\dfrac{ \rho \tpsize^{2}}{18\mu}$ is the particle relaxation time (s), $\rho$ is the density of the particle (kg m$^{-3}$), $\tpaccl = g \paccl $ is the particle acceleration (m s$^{-2}$) due to body forces, where $\paccl$ is the dimensionless unitary vector of the acceleration direction,
$\st=\dfrac{ \rho \refpsize^{2}}{18\mu} \dfrac{U}{L}$ is the Stokes number, and $\fr=\sqrt{\dfrac{U^2}{L g}}$ is the Froude number. The most common body forces considered is the gravity and therefore $\tpaccl$ is the gravity acceleration, but other body forces can be considered, e.g. electrostatic interactions. In this work we will also consider that the acceleration is the same (in magnitude and direction) for each particle.

\paragraph{Dilute limit:}
we assume the system is dilute, i.e., the overall volume fraction of the particles, which can be estimated as $F\refpsize^4$ (assuming the width of the distribution is of the same order of magnitude of the reference particle size), is large enough to have collisions between particles but small enough to assume particle momentum is independent from the concentration of particles and only dependent on particle Stokes number.

\paragraph{Particle diffusion:}
$\diff =\tdiff(\tpsize)/\diffref$ is the diffusivity ratio between particles of size $\tpsize$ and $\refpsize$. If we assume the Stokes-Einstein relation to compute the diffusion coefficient of spherical particles, i.e., 
$\tdiff(\refpsize)=\dfrac{\kappa_B T}{3 \pi\mu \refpsize}\,$, where $\kappa_B$ is the Boltzmann constant (kg m$^2$ s$^{-2}$ K$^{-1}$), and $T$ the temperature (K), then the diffusivity  ratio assume the simple form $\diff = \psize^{-1}$.

\paragraph{Particulate processes:}
$A(\tfe,\tfe,\tpsize)$, $B_1(\tfe,\tpsize)$, $B_2(\tfe,\tfe,\tpsize)$  are the dimensionless functions describing aggregation, single particle breakage, and two particles breakage respectively. In the literature on the PBE, the source term is often written in terms of death and birth rates, which both includes aggregation and breakage terms \citep{Ramkrishna2000}, e.g. an aggregation event can lead to the birth of a new particle of size $\tpsize$ or the death of a particle of that size. In this work, however, we are interested in analysing the source term in terms of particle events, for which the division in aggregation and breakage looks more clear. The description of the single or double particles events will be the main topic of next sections. We want to highlight here that the functions themselves are dimensionless, but they still depend on dimensional arguments (i.e. $\tfe$ and $\tpsize$). In the next section will rewrite these functions in terms of dimensionless arguments.  

\paragraph{Particle deposition:}
The boundary term $C(\tfe,\tpsize)$ represents the particle deposition on the surface of the porous matrix, described as a generic flux. This is an important process, highly dependent on particle size, but often neglected in PBE models and it appears as a boundary condition on $\Gamma$ in \cref{eq:dimpbe} (and accordingly in \cref{eq:adimpbe}) to take into account the modification of the PSD due to particle deposition on the surface of the porous matrix. In other words, the flux (convective and diffusive) of the particles normal to the interface, equals the ones deposited on the surface given by the deposition kinetic described through the function $C(\tfe,\tpsize)$.

In the following we will often indicate the processes of aggregation, breakage and surface deposition mechanisms with the collective name of \textit{reactions} even though no chemical reactions are involved. We decided for this extended use of the word \textit{reactions} to allow a more general mathematical treatment of the upscaling mechanisms by using effective \Dam number as the ratio between the reaction rate and the diffusive mass transport rate \citep{Bird2002}. 
The occurrence of this term in the following must be intended with this extended use in mind.

\section{Particulate processes}\label{sec:2events}

In this section we will perform a dimensional analysis of the different phenomena experienced by a particle. In turn, this analysis will allow us to properly define the source terms in \cref{eq:adimpbe} in terms of dimensionless quantities only. The most straightforward classifications of such events consider the number of particles involved. Here, we assume that events including more than three-particles are so rare that can be neglected. Therefore, in this analysis we will focus on one- and two-particles events. 

While the aggregation  depends on the collision frequency between two particles and the probability of the collision to be effective, for the breakage we can distinguish between single particle and two particles events instead. We call the former processes ``linear reactions'' and the latter ones ``non-linear reactions''. Breakup caused by interaction of particles can be due to impacts of particles \citep{chen2020collision,khalifa2021efficient} but also due to viscous shear stresses generated when particles come close \citep{Frungieri2021}.
Part of our analysis will include the dimensional analysis of the terms appearing inside the source components in \cref{eq:adimpbe} and rewriting these contributions as non-dimensional quantities. This, in turn, will lead to the definition of another non-dimensional number, an equivalent \Dam number, as the ratio between  reaction and diffusion time scales.

\subsection{Single-particle events: shear-induced interactions}
Single-particle events can be induced by different mechanism, such as local shear rate, or collisions with the walls.
Let us consider shear-induced events and a threshold force, which we call the \textit{critical} force, $\mathcal{F}_{cr}$, which represents the average  force needed to break the aggregate. Having described breakage events  in a statistical way, we can therefore express the breakage frequency $\tilde{\bo}$ as:
\begin{equation}\label{Eq:kernbreak}
    \tilde{\bo} = \tilde{B}^\prime_0 \cip{\frac{\mathcal{F}}{\mathcal{F}_{cr}}}^\gamma
\end{equation}
where $\mathcal{F}$ is the force acting on the aggregates and $\tilde{B^\prime}_0$ is the characteristic breakup frequency (s$^{-1}$) for the reference particle size. The parameter $\gamma \ge 0$ controls how likely is for a particle experiencing a force larger than the critical force to break. For large $\gamma$, the frequency of breakage below the critical force is negligible while it becomes very large for larger forces. In the limit of small $\gamma$, instead, the breakage is a constant independent random event. The law reported in \cref{Eq:kernbreak} is the same power law reported in the literature \citep{Barthelmes2003} having recognised the role played by the shear stress into this process. The critical force needed to break an aggregate should depend on the size of the aggregate itself. The evidence suggest that in the presence of shear big particles broke easily than small ones \citep{Serra1998}. We assume therefore for the critical force that $\mathcal{F}_{cr}=\mathcal{F}_0 \psize^{-\eta}$, where $\mathcal{F}_0$ is the reference force defined as the force needed to break a reference particle of size $\refpsize$ and $\eta>0$ is a parameter. Let us rewrite \cref{Eq:kernbreak} as:
\begin{align}\label{Eq:kernbreakadim}
    \tilde{\bo}
    &=
    \tilde{B}_0 \cip{\frac{\mu \tshear \tpsize^2}{\mathcal{F}_{0}\psize^\eta}}^\gamma
    =
    \tilde{B}_0  \cip{\frac{\mu U \refpsize^2}{L \mathcal{F}_{0}}}^\gamma \psize^{\gamma^\prime}\shear^\gamma  
    =
    \tilde{B}_0  \ndimbr \bo
\end{align}
where we have estimated the average force on the particle as $\mathcal{F} \propto \mu \tshear \tpsize^2$, proportional to viscous stresses and particle surface area, and where we have defined the new dimensionless number $\ndimbr = \cip{\frac{\mu U \refpsize^2}{L \mathcal{F}_{0}}}^\gamma$,  a dimensionless kernel
$
\beta = \psize^{\gamma^\prime}\shear^\gamma
$, and $\gamma^\prime = \gamma(\eta+2)$.

Another possibility, to have more consistent and comparable results for large different $\gamma$ and highly heterogeneous shear rates, we can normalise the shear by its gamma-moment, $\aver{\shear^\gamma}$ ( where $\aver{\cdot}$ is the average operator), resulting in the dimensionless kernel
\[
\beta = \psize^{\gamma^\prime}\frac{\shear^\gamma}{\aver{\shear^\gamma}}
\]
with constant $\tilde{B}_0\aver{\shear^\gamma}\ndimbr$ and effective \Dam number
$\Da{\beta} = \frac{\tilde{B}_0\aver{\shear^\gamma} L^2}{\diffref}\ndimbr$

\subsection{Two-Particle Events}\label{sec:2part}

When two-particle collide in a fluid, different mechanisms are involved, which can depend on the movement of the fluid itself. For this reason, two-particle collision events  are described as a sum of single mechanism, and the decisions on which ones are to be included depends on the particular problem we are considering. 

In this work we will focus on the Brownian, shear and settling velocity mechanism, even though others are available, such as the turbulent mechanism \citep{DiPasquale2012,Lavino2017}.
For each mechanism we define a constant, which we will indicate with $K_X$ with $X$ related to the particular mechanism considered, which collect all the dimensional reference quantities as well as the multiplicative coefficients for each mechanism. In turn, this constant divided by the reference volume will represents the \textit{characteristic frequency} of the particular mechanism (with dimension 1/s). 

For two-particle events, the \Dam number is defined as
$\Da{X} = K_X F\frac{L^2  \refpsize}{\diffref}$. $K_X$ is the process constant and its exact definition depends on the particular process considered. 
There will be therefore several \Dam numbers, one for each mechanism and they will be distinguished by using the same subscripts we used for the process constants. 

Assuming spherical particles of diameter $2\tpsize$, and considering only the rectilinear approximation of particle encounters\footnote{the more accurate curvilinear approximation can be used by adding a correction factor, typically particle size dependent, in front of the rectilinear collision frequency}, it is possible to discriminate which is the leading collision mechanism in the following way \citep{Serra1997}:

\begin{itemize}
\item Brownian motion for $\tpsize < 1 \mu \mathrm{m}$,  with a (dimensional) collision frequency between particle of diameters $\tpsize$ and $\tpsize_i$
\begin{equation}
	\tbrown(\tpsize,\tpsize_i)=2\pi\cip{\tpsize+\tpsize_i}\cip{\tdiff(\tpsize)+\tdiff(\tpsize_i)}
\end{equation}
In dimensionless form, assuming particles with constant density, this is equivalent to
\begin{equation}
	\tbrown(\tpsize,\tpsize_i)  = 2\pi\diffref\refpsize\cip{\psize+\psize_i}\sqp{\diff+\diff_i} =2\pi\diffref\refpsize\brown(\psize,\psize_i) 
	=\cfbrown\brown(\psize,\psize_i) 
\end{equation}
where we defined the Brownian constant $\cfbrown=\diffref2\pi\refpsize$.

\item Shear stress for $1\mu \mathrm{m} < \tpsize < 40\mu \mathrm{m}$, with
\begin{align*}
	\tkshear(\tpsize,\tpsize_i) & =\tfrac{1}{6}\del{\tpsize+\tpsize_i}^3 \tshear
\end{align*}

from which we obtain:
\begin{align}
	\tkshear(\tpsize,\tpsize_i)=\frac{U\refpsize^3 }{ L}\cip{\psize+\psize_i}^3 \shear \nonumber \\
	= \frac{U\refpsize^3 }{ L}\kshear(\psize,\psize_i) = \cfkshear \kshear(\psize,\psize_i) \,.
\end{align}
 We here defined the shear stress constant $\cfkshear=\frac{U\refpsize^3 }{ 6 L}$.


\item Differential settling for $\tpsize > 40\mu \mathrm{m}$, it can written as:
\begin{align}
	\tdiffsett(\tpsize,\tpsize_i) & =\dfrac{\pi}{4}\cip{\tpsize+\tpsize_i}^2 \left|\tilde{\vm}-\tilde{\vm}_i\right| 
\end{align}
and by considering \cref{eq:velpart}, we have:
\begin{align}
\tdiffsett(\tpsize,\tpsize_i) = & =\dfrac{\pi U  \refpsize^2}{4}\cip{\psize+\psize_i}^2 \left|\vm-\vm_i\right| \nonumber \\
	& = \dfrac{\pi U\st  \refpsize^2}{4\fr^2}  \cip{\psize+\psize_i}^2 \left|\paccl \psize^2 - \paccl \psize_i^2 \right| \nonumber \\
    & = \dfrac{\pi U\st  \refpsize^2}{4\fr^2}   \diffsett(\psize,\psize_i) \nonumber \\
	& = \cfdiffsett \diffsett(\psize,\psize_i)
\end{align}

with a constant equal to $\cfdiffsett=\dfrac{\pi\st U \refpsize^2}{4 \fr^2}$.

\end{itemize}

\subsection{Aggregation and breakage models}\label{sec:model}

%
In the literature on the PBE, two terms are usually defined to describe the aggregation and breakage, $\ao$ and $\bo$ which are known as the aggregation and breakage kernels (frequencies) respectively \citep{Marchisio2013}. However, in this work we want to present a different interpretation of the aggregation kernel as a term not related only to the aggregation but more in general as an interaction term between two particles (we remind here that we assumed that three or more particle events are negligible). The interaction between particles must be understood only as a collision and therefore in this work $\ao$ represents the collision kernel. 
In turn, this collision can create a bigger aggregate (aggregation) or breaks the particles into smaller ones (breakage).
In the literature, the breaking mechanism is usually described as one particle process, where the particle breaks either because of its interaction with the fluid flow in which it is immersed (e.g. Turbulent stresses, Viscous shear stresses, Shearing-off processes) \citep{Solsvik2013} or for its own interfacial instabilities \citep{Solsvik2013}. This description comes from the main application of the PBE in the past years, bubbles and drops. In this case, the only effect of the collision of two bubbles (or drops) is their coalescence into a new bubble (aggregation).
For (solid) particles the previous is not true. The collision between two particles can either create a bigger aggregate, if their structure allows them (e.g. for polymers \cite{DiPasquale2012,DiPasquale2013}), or they can break because of the collision, or nothing happen.
Therefore, in this formulation we will describe two-particle events leading to an aggregation, and two-particle events leading to a breakage.
Of course, there is still the possibility that particles collide without aggregation or breaking.

In order to distinguish among these three possibilities (aggregation, breakdown, or nothing) we define two new parameters $\kappa_A$ and $\kappa_B$, the former represents the probability that a collision leads to aggregation, the latter is the probability that a collision event leads to  breakage. 
We note that $\kappa_B$ and $\kappa_A$ can be considered as the efficiency of aggregation and breakage respectively. Being defined as efficiencies then $0\leq \kappa_A \leq 1$, and the same for $\kappa_B$. We can distinguish two possible scenarios:
\begin{itemize}
    \item if $ \kappa_B + \kappa_A = 1$, then we assume that each collision event is going to lead either to an aggregation or to a breakage
    \item if $\kappa_B + \kappa_A < 1$, a collision event has one result among these three events: an aggregation, a breakage or nothing. 
\end{itemize}
It can be easily seen that each possible special cases is included by choosing particular values for $\kappa_A$ and $\kappa_B$, e.g., if we assume $\kappa_A=1$ then each and every collision will result in an aggregation (see e.g. \citep{DiPasquale2012}).

We start our analysis by considering the aggregation $A(\tfe,\tfe,\tpsize)$ term in \cref{eq:adimpbe}. This latter object can be considered a functional with respect the PSD, and a function with respect the particle size $\psize$. It can be described as follow, where we note that the dependence on time and space variables have been dropped for the sake of legibility \citep{Vanni2000,Serra1998}:
\begin{equation}\label{eq:A}
A(\tfe,\tfe,\tpsize) = \sum_{X}A^-_X(\tfe,\tfe,\tpsize) + \sum_{X}A^+_X(\tfe,\tfe,\tpsize) 
\end{equation}
here, the sums run over the different mechanisms described in \cref{sec:2events}, i.e. $X \in \{B,S,D\}$.
The two terms in \cref{eq:A} represent respectively the creation of particles of size $\tpsize$, and the disappearing of particle of size $\tpsize$ both following an aggregation event.

Writing explicitly the first term, including the adimensionalisation constant $\frac{L^2}{\diffref F}$ (see \cref{eq:adimpbe}), we have: 
\begin{align}\label{Eq:Amin}
    \frac{L^2}{\tdiff F} \tilde{A}_X^-(\tfe,\tfe,\tpsize) 
    & = -\frac {1}{2}\frac{L^2}{\diffref F}  \int_0^{+\infty}\kappa_A\tilde{\ao}_X(\tpsize,\tpsize_i) \, \tfe \tfe_i \de \tpsize_i \nonumber \\
    & =- \frac {1}{2}\frac{L^2}{\diffref F}  \int_0^{+\infty}\kappa_A K_X \ao_X(\psize,\psize_i) \, F \fe F \fe_i \refpsize \de \psize_i \nonumber \\
    & = - \frac {1}{2}\Da{X} \fe \int_0^{+\infty}\kappa_A\ao_X(\psize,\psize_i) \, \fe_i \de \psize_i = \Da{X} A_X^-(\fe,\fe_i,\psize). 
\end{align}
where $\ao_X$ represents one of the interaction kernels defined above, and the subscript $X$ refers to the fact that one or more of those mechanism can be included, depending on the particular problem considered, $\Da{X}$ is the \Dam number for the aggregation referring to the particular process $X$, $\kappa_A$ is the probability that a collision leads to an aggregation and $\tfe_i$ is a shorthand notation for $\tfe(\tpsize_i)$. $A^-(\tfe,\tfe,\tpsize)$ represents the reduction of the population of particles of size $\psize$ due to their aggregation with any other particle. 
For the second term in \cref{eq:A} we can write, with a derivation similar to the one shown in \cref{Eq:Amin}:
\begin{equation}\label{Eq:Aplus}
    A_X^+(\tfe,\tfe,\tpsize) =  \Da{X} A_X^+(\fe,\fe,\psize) = \Da{X}\frac {1}{2} \int_0^\psize \kappa_A \ao_X(\psize,\psize-\psize_i) \, \fe(\psize_i)\, \fe(\psize-\psize_i)\de \psize_i \,.
\end{equation}
$A^+(\tfe,\tfe,\tpsize)$ represents the increase in the number of particle of size $\psize$ due to aggregation of a particle of size $\psize_i < \psize$ and $\psize-\psize_i$.

In the case of breaking, we have two different terms to consider $\tilde{B}_1$ and $\tilde{B}_2$ representing the single particle and the two particle events respectively. 
As we did for the aggregation case, we need to distinguish between two different occurrences in which a new particle of size $\psize$ is created by a breakage, or a particle of size $\psize$ is destroyed by the breakage. However, the breakage mechanism, differently from the aggregation one, contains a further term in its formulation, the so-called daughter distribution function, $\tDdistr(\tpsize_i,\tpsize)$. When a particle with size $\tpsize_i > \tpsize$ breaks it can produce a range of fragments (with the obvious constraints that no fragment can be greater than the original one). In our formulation we need to keep track of this effect to include in our count all the new fragments with size $\psize$ created. It turns out that it is convenient to redefine the daughter distribution function  $\tDdistr(\tpsize_i,\tpsize)$ as:
\begin{align}\label{eq:Ddistr}
    \tGDdistr(\tpsize_i,\tpsize) =  \tDdistr(\tpsize_i,\tpsize) - \tilde{\delta}(\tpsize_i-\tpsize)
\end{align}
where $\tilde{\delta}$ is the Dirac delta-function. We can observe that the new generalised daughter distribution function $\tGDdistr(\tpsize_i,\tpsize) $ has the right measures of the inverse of a length. The two terms in \cref{eq:Ddistr} take into account the fact that the breaking of a particle of size $\tpsize_i> \tpsize$ can increase the population of particles with size $\tpsize$, whereas the breaking of a particle with size $\tpsize$ always decreases the population of such particles.

The single particle events $B_1$ is now given by:
\begin{align}\label{eq:breakI}
     \frac{L^2}{\diffref F}\tilde{B}_1(\tfe,\tpsize) &
    = \frac{L^2}{\diffref F}\int_{\tpsize}^{+\infty} \tGDdistr(\tpsize_i,\tpsize) \, \tilde{\bo}(\tpsize_i)\, \tfe(\tpsize_i) \de \tpsize_i  \nonumber \\
     & = \frac{\tilde{B}_0\aver{\shear^\gamma} L^2}{\diffref}\ndimbr  \int_{\psize}^{+\infty} \GDdistr(\psize_i,\psize)\, \bo(\psize_i)\, \fe(\psize_i) \de \psize_i   = \Da{\beta} B_1(\fe,\psize).
\end{align}
where $\GDdistr(\psize_i,\psize) = \refpsize \tGDdistr(\psize_i,\psize)$, and we included the non-dimensional number $\ndimbr$ in the definition of $\Da{\beta}$.

The formulation of two-particle events for breakage is different from the ones shown for aggregation, because here either of the two particles involved in the collision can break, leading to:

\begin{subequations}\label{eq:explB}
\begin{align}
\tilde{B}_X(\tfe,\tfe,\tpsize) = & \frac {1}{2}\int_{\tpsize}^{+\infty}\int_0^{+\infty} \kappa_B\cip{\tGDdistr(\tpsize_i,\tpsize)+\tGDdistr(\tpsize_j,\tpsize)}\, \tilde{\ao}_X(\tpsize_i,\tpsize_j) \tfe(\tpsize_i)\tfe(\tpsize_j) \de \tpsize_i \de \tpsize_j  \label{eq:exB1} \,.
\end{align}
\end{subequations}
where $X$ represents the collision mechanism, we have assumed that each particle breakup event is independent and the breakup frequency $\kappa_B$ is constant with respect to the particle size. 
Its non-dimensional versions is given by:
\begin{align}\label{eq:breakII}
    \frac{L^2}{\diffref F}\tilde{B}_X(\tfe,\tfe,\tpsize) & = \frac {1}{2} \frac{L^2}{\diffref F}\int_{\psize}^{+\infty}\int_0^{+\infty} \kappa_B\frac{1}{\refpsize}\cip{\GDdistr(\tpsize_i,\tpsize)+\GDdistr(\tpsize_j,\tpsize)}\, K_X\ao_X(\psize_i,\psize_j) F\tfe(\psize_i)F\fe(\tpsize_j) \refpsize \de \psize_i \refpsize \de \psize_j \nonumber \\
    & = \frac{L^2K_X F \refpsize}{\diffref}\int_{\psize}^{+\infty}\int_0^{+\infty} \kappa_B\cip{\GDdistr(\psize_i,\psize)+\GDdistr(\psize_j,\psize)}\,\ao_X(\psize_i,\psize_j) \tfe(\psize_i)\fe(\tpsize_j) \de \psize_i \de \psize_j\nonumber \\
     & = \frac{L^2K_X F \refpsize}{\diffref} B_X(\fe,\fe,\psize) = \Da{X}  B_X(\fe,\fe,\psize) \,.
\end{align} 
We note here an important symmetry between aggregation and two-particle events breaking, the \Dam number is the same for both of them, its only dependence is on the particular two-particle collision mechanism (represented by the subscript $X$). 

Now we are in the position to show explicitly  the \Dam number  for the different mechanism in \cref{sec:2part}. 
We obtain for each mechanism:
\begin{itemize}
    \item Brownian motion: $\Da{B} =  2\pi \refpsize \diffref \dfrac{F L^2 \refpsize}{\diffref }  = 2\pi F \cip{\refpsize L}^2 $
    \item Shear stress: $\Da{S} = \dfrac{U\refpsize^3 }{ 6 L} \dfrac{F L^2 \refpsize}{\diffref }  = \dfrac{1}{6}F\pe\, \refpsize^4$ 
    \item Differential Settling: $\Da{D} = \dfrac{\pi\st U \refpsize^2}{4 \fr^2} \dfrac{F L^2 \refpsize}{\diffref } = \dfrac{\pi}{4} \dfrac{\st  \pe}{\fr^2} F  \refpsize^3L$ 
\end{itemize}
\Cref{table} summarises this convenient dimensionless formulation,   and will allow us to significantly simplify the parametric sweeping and analysis of the macroscopic rate of collision/breakage events in the next section.

\begin{table}[htp]
\begin{center}
\begin{tabular}{p{3cm}| c c c c}
       Particulate process, $X$  & Frequency & Characteristic & Damk{\"o}hler & Damk{\"o}hler \\
         & kernel & frequency &  number (diffusive) &  number (convective) \\
        &  & $K$ & $\Da{X}$ & $\Da{X}^{II}$ \\ 
        \midrule
& \multicolumn{4}{c}{Two-Particle Events (aggregation or breakage)} \\
      Brownian motion, B & $\cip{\psize_j+\psize_i}\sqp{\diff_j+\diff_i}$ &
      $2\pi \diffref \refpsize$ & $2\pi F \cip{\refpsize L}^2$ & $2\pi F \dfrac{\diffref}{U} \refpsize^2 L$  \\ 
      Shear, S & $\left(\psize_i+\psize_j\right)^3 \sigma$ & $\dfrac 16 \dfrac{\refpsize^3 U}{L}$ & $\dfrac{1}{6}F\pe\, \refpsize^4$   & $\dfrac{1}{6}F\, \refpsize^4$  \\
      Differential settling, D & $\cip{\psize_i+\psize_j}^2 |\psize_{i}^2-\psize_j^2|$ & $\dfrac{\pi}{4}\dfrac{\refpsize^2 U\st}{\fr^2}$ & $\dfrac{\pi}{4} \dfrac{\st  \pe}{\fr^2} F  \refpsize^3L$ & $\dfrac{\pi}{4} \dfrac{\st }{\fr^2} F  \refpsize^3L$ \\
\midrule
& \multicolumn{4}{c}{Single-Particle Events (breakage)} \\
      Shear, $\ndimbr = \cip{\frac{\mu U \refpsize^2}{L \mathcal{F}_{0}}}^\gamma$ & $\psize^{\gamma^\prime}\frac{\shear^\gamma}{\aver{\shear^\gamma}} $ & $\tilde{B}_0\aver{\shear^\gamma}\ndimbr$ & $\dfrac{\tilde{B}_0\aver{\shear^\gamma} L^2}{\diffref}\ndimbr$ & $\dfrac{\tilde{B}_0\aver{\shear^\gamma} L}{U}\ndimbr$ \\
  \end{tabular}
  \end{center}
  \caption{Definition of the dimensionless relevant quantity for the study of the  breakage and aggregation problem}
\label{table}
\end{table}%
%
%

\subsection{Particle deposition}\label{sec:deposition}

We provide here a short description of the particle deposition mechanism $C$ in \cref{eq:finalpbe}. A more detailed analysis of this term and its interactions with the aggregation/breakage processes will be subject of future works.

The deposition term $C$ in \cref{eq:finalpbe}, is a simple, yet general, way of incorporating several deposition mechanisms into a single (effective, meso-scale) boundary condition, relating the flux at the boundary with a deposition kinetics. It is common to include in the fluxes only the diffusion term (see e.g., \citep{Grigoriev2020,Boccardo2017}, assuming therefore a zero velocity of the particles at the wall. This is true only for solute or very small particles, while, as we have seen in  \cref{eq:velpart}, inertial particles can have a non-zero velocity at the boundary due to, for example gravity or electrostatic attractions. It is recommended, therefore, to write a more general boundary condition including also the convective fluxes.

We assume that the deposition mechanism can be described as a first-order, possibly non-linear, irreversible kinetics:
\begin{equation}\label{eq:C}
   \tilde{C} = \tilde{k}_c(\tpsize) \tfe
\end{equation}
where $\tilde{k}_c(\tpsize)<0$ is the deposition constant as a function of the particle size. For example, steric deposition (deposition due to the finite size of the particles colliding into the solid) can be estimated to be $\propto r^a$ with $a\le 1$. Other deposition mechanisms (chemically or electrostatic-induced) might have in general a positive or negative $a$. It is important  to notice that we consider here only deposition as irreversible so detachment is not contemplated.

In the case of no deposition $\tilde{k}_c=0$ and \cref{eq:pbeG} represents simply a (Robin- or Neumann-type) boundary condition stating that the total flux of particles onto the boundary is zero. In the case of infinitely fast (i.e., perfect) deposition, each particle arriving on the boundary deposits instantaneously and are no longer in the system. This means that there are no "free" particles on $\Gamma$ and we obtain the homogeneous Dirichlet boundary condition $\tfe = 0$ on $\Gamma$. Simple extensions for reversible linear deposition are also possible by adding an additional balance equation for deposited particles and will be explored in further work.

The non-dimensional version of \cref{eq:pbeG}  can be written again as:
\begin{equation}
   C = \Da{c} k_c(\psize) \fe
\end{equation}
where $k_c(\psize)$ is the non-dimensional deposition constant rate, and  $\Da{c}$ is the reaction coefficient for the deposition events, (\Dam number for the deposition). The explicit definition of $\Da{c}$  depends on the specific deposition mechanism and will be considered in details in future publications.


\section{Upscaling}

The final form of the dimensionless PBE derived in the previous section reads as follows:

\begin{align}\label{eq:finalpbe}
\pard{f}{t}  &+  \pe  \nabla \cdot \cip{\fe\vm}  -   \diff \Delta \fe  = \\
&\sum_{X} \left[ \Da{X}\frac {1}{2} \int_0^\psize \kappa_A \ao_X(\psize,\psize-\psize_i) \, \fe(\psize_i)\, \fe(\psize-\psize_i)\de \psize_i - \frac {1}{2}\Da{X} \fe \int_0^{+\infty}\kappa_A\ao_X(\psize,\psize_i)  \, \fe_i \de \psize_i \right. \nonumber \\
&  \left. + \int_0^{+\infty} \kappa_B\cip{\GDdistr(\psize_i,\psize)+\GDdistr(\psize_j,\psize)}\,\ao_X(\psize_i,\psize_j) \tfe(\psize_i)\fe(\tpsize_j) \de \psize_i \de \psize_j\right]  + \Da{\beta}\int_{\psize}^{+\infty} \GDdistr(\psize_i,\psize)\, \bo(\psize_i)\, \fe(\psize_i) \de \psize_i \nonumber 
\end{align}
with boundary conditions
\[
\nb\cdot\del{\pe\, \vm\fe -  \diff \grad \fe} =  \Da{c} k_c(\psize) \fe \,\,.
\]

The detailed analysis and resolution of the different terms (aggregation, breakage, and deposition) in \cref{eq:finalpbe} depends critically on the specific system, and in particular, on their dynamics with respect the typical time scales of the evolution of the flow in the system under study. In order to quantify these differences, we can use the \Dam and the $\pe$, and we can distinguish three different regimes. \\

When $\da\ll\pe$, we can assume that the breakage, aggregation and deposition processes are slower than mixing. This latter fact, in turn, means that we can safely neglect both the interactions between these processes and the secondary collision events. The upscaled reactions can be trivially transferred from micro- to macro-scale as they are not mixing-limited. This is the case studied with periodic homogenisation by \citet{Krehel2015} and it can be shown that the same applies also to non-linear rates.\\

When $\da\approx\pe$, the effective reactions (i.e. microscopically caused by average frequency of particle encounters) start to be significantly affected by advection patterns and influence effective upscaled velocity and dispersion. In particular, the effective reaction rate is no more linearly dependent on the microscopic  reaction/collision rate. Non-linearities, in general, can be upscaled only locally. Furthermore, poly-dispersity starts to play an important role, as successive collisions can happen within the same upscaling averaging volume.\\

Finally, when $\da\to\infty$ mixing-limited asymptotic behaviours can be identified for the reaction rate. When the reaction happens only on a sub-domain or lower dimensional manifold, this saturates to a P\'eclet-dependent constant reaction rate \citep{Boccardo2017,fastreaction}, while, if the reaction (although heterogenous) happens in the whole domain, we expect the effective reaction rate to grow to infinity as it is limited but not completely impeded by the incomplete mixing.

Applying this principle to the PBE \cref{eq:finalpbe}, and using the dimensionless numbers in \cref{table}, we can find the conditions that ensure the system is well mixed, and therefore processes (including the non-linear ones) can be decoupled and upscaled with a effective macroscopic constant, proportional to the pore-scale one. The conditions on the reference number density and particle size are
\begin{align}
F\refpsize^2 &< \frac{U}{2\pi\diffref L}\,, \nonumber \\
F\refpsize^4 &< 6\,, \nonumber \\
F\refpsize^3 &< \frac{4\fr^2}{\pi\st L}\,, \nonumber \\
\refpsize^2 &< \frac{L \mathcal{F}_{0}}{\mu U}\left( \frac{U}{L\tilde B_0} \right)^{1/\gamma}\,. \nonumber
\end{align}
It is important to recall that, although the constraint for shear-induced breakage does not explicitly involve the reference number density $F$, we have assumed that the system is dilute. The second constraint, coming from shear-induced collisions is similar, as expected, to the dilute limit approximation.

In case of wide particle size distributions, where the minimum and maximum particle size are orders of magnitude different from the reference particle size, the distinction above is not valid for all particles and a robust upscaling procedure, valid for all cases, should be considered. This will be the aim of our future works.

\subsection{Formal averaging and closure problems}\label{sec:homog}
Let us define the following (static) volume averaging operators:
\begin{equation}\label{eq:medie}
\mV{\cdot} = \dfrac{1}{|\por|} \int_{\por} \cdot \textrm{ d}v\,,			\quad	\quad	\mG{\cdot} = \dfrac{1}{|\Gamma|} \int_\Gamma \cdot \textrm{ d}s
\end{equation}
%

Instead of the standard volume averaging approach \citep{whitaker1998}, that involves either a moving average or a local kernel, since we are now only interested in the upscaled particulate processes, we can simply apply this static volume averaging to
\eq{finalpbe}, obtaining:
\begin{align}\label{eq:pbeav}
    &\pa_t\aver{ \fe} + \aver{\pe \nabla \cdot \left(\vm\fe\right) -  \diff \Delta \fe} = \aver{B_1\cip{\fe,\psize}} + \sum_{X}
\Da{X}\cip{\aver{A_X^+\cip{\fe,\fe,\psize}}+ \aver{A_X^-\cip{\fe,\fe,\psize}} +  \aver{B_X\cip{\fe,\fe,\psize}}}
\end{align}
%
%
%
where we have to remember that $\fe$ and all the kernels are functions of space and particle size, and therefore they all give rise to closure problems.


Applying the divergence theorem, the second and third terms on the lhs of \cref{eq:pbeav} can be rewritten as:
\begin{align}\label{eq:up2}
\mV{\pe\nabla \cdot \left( \vm f \right) - \diff \Delta f} &= \mV{\nabla \cdot \left( \pe \, \vm f -  \diff  \nabla f\right) } = \nonumber \\
&=\frac{|\Gamma|}{|\por|} \mG{\nb\cdot\cip{\pe\, \vm\fe -  \diff \grad \fe}}
+
\frac{1}{|\por|} \int_{\partial\por} \bm{n}\cdot \left(\pe \, \vm f - \diff \nabla f \right) \textrm{ d}s
= \nonumber \\
&=\frac{|\Gamma|}{|\por|} \mG{ C(\fe,\delta)}
+
\frac{1}{|\por|} \mathfrak{F}
\end{align}

where $\frac{1}{|\por|}$ is the pore specific surface area, $\partial \por$ are the external REV boundaries and $\mathfrak{F}$ are the total fluxes through them.
The upscaling problem, assuming a macroscopic equilibrium (steady state) configurations, consists in finding an approximation for
\[
\mathfrak{F}
=
\aver{B_1\cip{\fe,\psize}} + \sum_{X}
\Da{X}\cip{\aver{A_X^+\cip{\fe,\fe,\psize}}+ \aver{A_X^-\cip{\fe,\fe,\psize}} +  \aver{B_X\cip{\fe,\fe,\psize}}}
-
\frac{|\Gamma|}{|\por|} \mG{ C(\fe,\delta)}
\]
as a function of macroscopic variable $\aver{\fe}$ and appropriate macroscopic effective parameters. This is tackled in homogenisation \citep{Hornung} and volume averaging theory \citep{whitaker1998} by expanding the variable $\fe$ as $\fe=\phi\aver{\fe} + \phi \mathbf{w}\cdot\nabla\fe$, where $\phi$ is a function needed to decouple fast reactions \citep{fastreaction}, neglected (it tends to one) in the standard upscaling for slow reactions, and $\mathbf{w}$ is given by a local cell problem. This approach, however, in case of non-linear terms, leads to $\phi$, $\aver{\fe}$ and $\mathbf{w}$ being coupled.

\subsection{Upscaling of particle linear processes}
\label{sec:upscbeta}

\noindent In this section we specialise our discussion to shear-induced linear events only, leaving the detailed analysis of the other terms in \cref{eq:pbeav} for future works.  Among all the possible events described in \cref{sec:model}, the only linear ones are the deposition (when $k_c$ does not depend on $\fe$) and the single-particle breakage (see \cref{eq:breakI}).

To avoid solving the PDE in the five-dimensional space, we focus on steady state solutions and on a single particle size. In general, however, a breakage of a particle of size $\psize$ generates smaller particles and these have a finite probability of undergoing another breakage event. This results in a one-way coupled cascade of equations where, to solve for particles of size $\psize$ we first need to solve for all larger particles.
However, we can neglect this coupling if one of the following  assumptions hold:
\begin{itemize}
\item the initial distribution (i.e., the inlet distribution for steady state problems) is concentrated around a single particle size which, having neglected aggregation, represents the largest possible size which is therefore fully decoupled from the others
\item the breakage frequencies are small enough to assume that, in a single REV, each particle can undergo only one breakage event during its residence time
\item the daughter distribution function $\mathcal{A}=0$, i.e., particles are removed from the system (or no longer tracked) once they break. This is a common approach for estimating the frequency of shear-induced events without tracking the subsequent particle dynamics \citep{saha2016breakup}.
\end{itemize}

Therefore, to upscale the breakup due to shear with a bottom up approach, we fix a single particle size and we can then study the behaviour of a \textit{mono-dispersed} sample of particles. Although it is important to stress that the behaviour of a poly-dispersed population is not the trivial superimposition of mono-dispersed particle dynamics, this assumption reduces the complexity of the system allowing us to focus on the role of shear heterogeneity on breakage phenomena.
We therefore consider the simplified version of \cref{eq:finalpbe}
\begin{equation}\label{eq:redpbe}
\dfrac{\partial f}{\partial t} + \pe\ \nabla \cdot \left(\vm f \right) - \diff \Delta f =
-\Da{\beta} \beta f\ ,
\end{equation}
where $f$ can now be interpreted as the concentration of particles of size $\psize$.

The upscaling of deposition phenomena is a widely studied problem \citep{Boccardo2017,Elimelech1995,Krehel2015,Boccardo2019}, we assume here Neumann no-flux boundary conditions
\begin{equation}\label{eq:eb2diff_BC}
\nb\cdot\cip{\pe\, \vm\fe -  \diff \grad \fe} = 0 \ .
\end{equation}
It is important to notice, however, that, when the assumptions above are relaxed, particles smaller than $\psize$ will have an extra source term coming from breakage of larger particles. As we will see later, this is an heterogenous term, which will further generate heterogeneous spatial distributions, affecting significantly deposition but also dispersion properties.

Applying the averaging described above to  \cref{eq:eb2diff_BC,eq:redpbe} we seek a macroscopic representation of the type
\begin{equation}\label{eq:DAM}
\mathfrak{F}=-  \aver{\Da{\beta}\beta \fe } = -\aver{\Da{mi}\fe}  \approx -  \Da{Ma}\aver{\fe} =  \Da{\beta,Ma}\beta_{\mathrm{Ma}}\aver{\fe}\,  ,
\end{equation}
%
%
%
where $\beta$, as defined in \cref{Eq:kernbreak}, is the breakage kernel and depends on shear (and therefore space $\xb$) and particle size $\psize$.
Having fixed here the size $\psize$, we can include it in an overall pore-scale reaction constant, $\Da{\mathrm{mi}}=\Da{\beta}\, \beta(\psize)$.
We have assumed here the existence of an ``effective'' macro-scale reaction rate $\Da{Ma}$ which describes the effects observed at the macro-scale obtained by this particular mechanism.
The macroscopic reaction constant can then be split into the product of a macroscopic breakage kernel $\beta_{\mathrm{Ma}}$, which carries all the dependencies of particle size (and is one for the reference particle of size $\refpsize$), and a breakage macroscopic constant $\Da{\beta,Ma}$.

While in this case it is possible to evaluate directly the term $\mV{\beta f}$, it is generally more convenient to compute the inlet and outlet fluxes of \cref{eq:redpbe}, as this would work also in presence of more complex (reactive) boundary conditions.
The macroscopic reaction constant $\Da{\mathrm{Ma}}$, having fixed a velocity field $\vm$ and a reaction kernel $\beta$, is therefore a function of only two terms: $\frac{\pe}{\diff}$ and $\frac{\Da{\beta}\, \beta}{\diff}$, which can be directly linked to the particle size $\psize$.

From the practical point of view, \cref{eq:DAM} requires the (non-trivial for complex pore geometries) stationary solution of \eq{redpbe} which exists only when a concentration gradient is imposed at the boundaries. The assumption of constant concentration in the inlet is however not a good representation of the asymptotic REV, so quasi-periodic BCs can be used \citep{Boccardo2018,Boccardo2017}. Then, volumetric and surface integrals have to be extracted. Both steps might introduce cancellation and numerical errors due to the fact that, when the reaction rate is high, the average $\aver{\fe}$ vanishes. A convenient alternative is to compute the effective reaction rate $\Da{\mathrm{Ma}}$ as the first smallest eigenvalue of the differential operator \citep{Allaire2007,Mauri1991,Municchi2020}, under periodic (or Neumann) external boundary conditions. The corresponding eigenfunction has, in fact, the physical meaning of the leading order self-similar solution (e.g. solutions that are equals up to a constant) of the equation. The eigenvalue is instead the required multiplicative constant that represents the macroscopic effective reaction rate. Our future works will explore more in details this approach for various physical problems. This is used here only as a comparison for the channel flow example in the next section.

%
%

Two-particle processes could be similarly studied in the assumption that the second particle $j$ has a fixed size $\psize_{j}$ and its concentration field is constant. This can be also thought as a linearisation, assuming that the concentration of particle of size $\psize_{j}$ changes at a much slower speed. A proper non-linear generalisation and perturbation approaches will be studied in future works.

\section{Numerical examples}

In this section we apply the framework derived previously (in particular in \cref{sec:upscbeta}) to different example problems to show the potentiality of the whole approach.

\begin{figure}[htbp]
\label{fig:geometries}
\centering
\includegraphics[width=.3\textwidth]{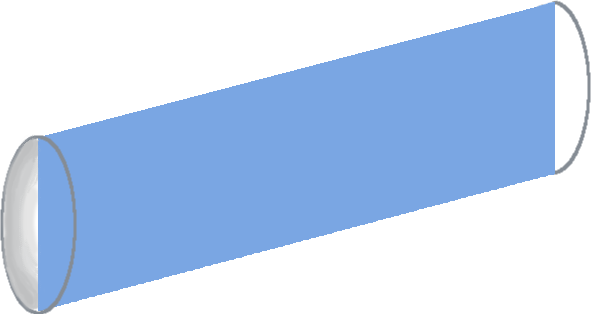}\quad
\includegraphics[width=.45\textwidth]{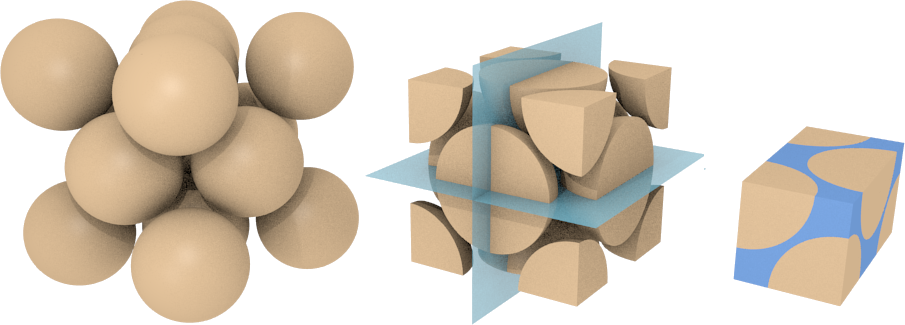}
\caption{Sketch of the two-dimensional channel (figure on the far-left) and FCC packing domains with shown the cut performed to obtain the final domain used in the simulation (figure on the far-right).}
\end{figure}

Two different geometries were chosen: a two-dimensional channel (CH) with an aspect ratio of 5 and a three-dimensional face-centered cubic domain (FCC) with porosity of $40\%$, as depicted in \fig{geometries}. A Cartesian orthonormal reference frame is used, both for the CH geometry, $\{O,\mathbf{e}_x ,\mathbf{e}_y \}$, and the FCC geometry, $\{O,\mathbf{e}_x, \mathbf{e}_y, \mathbf{e}_z\}$, where $O$ is the origin and $\mathbf{e}_x$ is the unit vector parallel to the main flow direction.
Given the symmetry of the FCC packing \citep{Crevacore2016,Boccardo2017}, just one fourth of the cubic structure was discretised and used for simulation purposes.
All simulations are performed with the OpenFOAM\textregistered finite-volumes open-source library

For each geometry, first Stokes equation (\eqs{campoDiMoto}) is solved  assuming periodic boundary conditions on the external boundaries of the domain and no slip condition on the solid walls, i.e., the lateral boundaries of the CH, and on the grain surface of the FCC. To drive the flow, a pressure drop between inlet and outlet boundaries, such that the resulting velocity has a unitary spatial average, i.e., $\mV{\vm}=1$, in both geometries.

The solution of the PBE transport equation, with the hypothesis considered here,  \cref{eq:redpbe},  reduces to the solution of a simple linear advection diffusion equation with appropriate heterogeneous reaction terms.
In this work, we present a parametric analysis obtained by considering different values of the $\gamma$ exponent in the breakage kernel (see \cref{Eq:kernbreak,Eq:kernbreakadim}), and of the  \peclet number. In particular, for the parameter $\gamma$ we have considered:
\begin{itemize}
\item $\gamma = 1$ to study the linear dependence of the reaction coefficient on the shear rate;
\item $\gamma = 2,3,10$ to investigate a power law of the share rate.
\end{itemize}
Whereas for the \peclet number, in the form of overall \peclet number: $\pe/\diff$, we considered a range of values between \Die{-2} and \Die{2}. In the rest of this work, in order to use a lighter notation, we will use the symbol $\pe$ to indicate the overall  \peclet number. All the results shown in the next sections must be intended with the diffusion coefficient implied in the definition of the $\pe$. We recall, in fact, that, even in the dimensionless formulation, particles of different sizes can have different diffusion coefficients as well as different reaction constant (breakage kernels).

Finally, for each combination of $\pe$ and $\gamma$, we assigned a broad range of values for the breakup reaction constant $\Da{mi}$ (which was defined in \eq{DAM}). 
This dimensionless formulation  allows us to run the whole set of simulations varying only three parameters $\gamma,\pe,\Da{mi}$ and subsequently remap the results and study the dependencies on the internal parameters $\psize,\pe, D$ and the other dimensionless groups appearing in the \Dam numbers (see \cref{table}). 
After reaching steady state, integration \cref{eq:DAM} gives a macro-scale effective reaction rate as a function of the above parameters. More details of the computational setup and numerical schemes are reported in previous works \citep{Boccardo2020}.

\subsection{The two-dimensional channel} \label{Sec:channel}

\begin{figure}[htb]
\centering
\includegraphics[width=9cm, height=6.5cm]{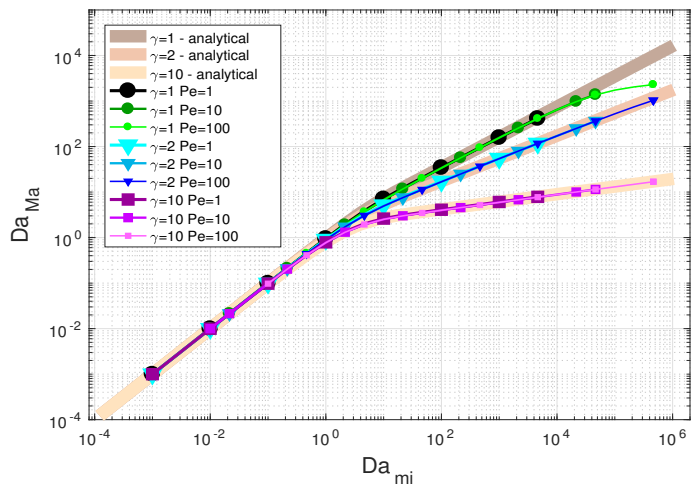}
\caption{Two-dimensional channel: macro- vs. micro-scale \Dam numbers for different $\gamma$ and $\pe$. We also report as comparison the analytical law extracted in \cref{Eq:damm,Eq:damM,} for the same values of $\gamma$.}
\label{fig:channel}
\end{figure}

We start our analysis of the results by reporting in \cref{fig:channel} the macroscopic \Dam number as function of the microscopic one, for all the different operating conditions we considered in this work. 

Under the chosen parametrisation, two different trends can be highlighted. For small values of $\Da{mi}$, to which correspond small frequencies $\beta_0$, the equivalent macroscopic reaction rates $\mathrm{Da}_{\mathrm{Ma}} \,$ collapse on a single curve due to the dimensional scaling. This is due to the slow reaction kinetics that, although heterogeneous, is quickly homogenised by diffusion. On the other and, at higher breakage frequencies, (i.e. high values of $\Da{mi}$)  a strong dependence on the shear exponent $\gamma$ appears evident.

For the two-dimensional channel we can observe that there is no dependence on the \peclet number of the results. This is due to the pathological nature of the channel flow where mixing is purely due to lateral diffusion. However, despite being a simplified case, channel flows are the phenomenological basis for more complex porous media mixing structures \citep{dentz2018mechanisms}.

From the results shown in \cref{fig:channel}, the dependence of the macroscopic breakup rate is well described by:
\begin{equation}\label{Eq:damm}
\mathrm{Da}_{\mathrm{Ma}} \sim \mathrm{Da}_{\mathrm{mi}}\, , \quad \quad \mathrm{for}\ \mathrm{Da}_{\mathrm{mi}}<1 \, ,
\end{equation}
\begin{equation}\label{Eq:damM}
\mathrm{Da}_{\mathrm{Ma}} \sim \mathrm{Da}_{\mathrm{mi}}^{\frac{2}{\gamma+2}} \, , \quad \quad \mathrm{for}\ \mathrm{Da}_{\mathrm{mi}}>1 \, ,
\end{equation}
where the dependence on the particle size is implicitly taken into account by $\mathrm{Da}_{\mathrm{mi}}$ by its dependence on the breakage kernel (see \cref{eq:DAM}) and on the diffusion coefficient. This latter effect is compatible with the fact that, when reaction events (i.e. the breakage events) are faster then diffusion, the system enters in a mixing-limited regime where a non-linear dependence of the macro-scale reaction on the micro-scale breakage frequency is expected. However, contrarily to surface reactions \citep{Boccardo2017,fastreaction}
This dependence is thus well characterised as a power-law, where the controlling parameter is the shear rate exponent $\gamma$.

\begin{figure}[htb]
\label{fig:channel-profiles}
\centering
\includegraphics[width=9cm, height=6cm]{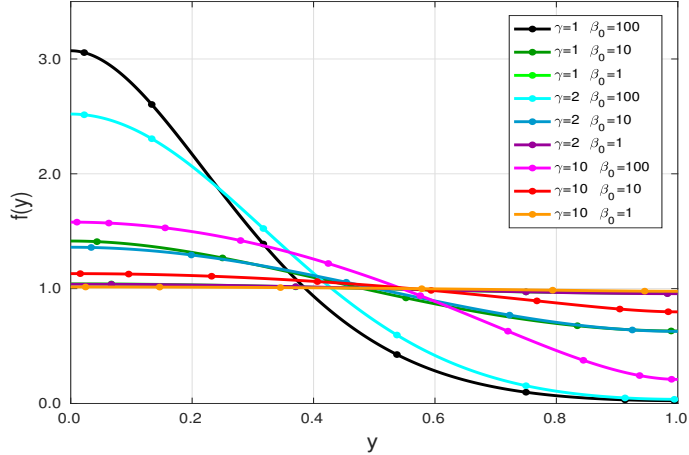}
\caption{Semi-analytical stationary concentration profiles for the Two-dimensional channel or different $\gamma$ and $\pe$.}
\end{figure}

This evidence is confirmed, in turn, by semi-analytical reference results. These are obtained on the reference domain $[0,1]$ (i.e. the channel cross-section), extracting the effective reaction rate by computing the principal eigenvalue of the Generalised Airy Equation. This simple 1D elliptic equation is obtained assuming a periodic stationary profile
\begin{equation}
\begin{aligned}\label{Eq:analyt}
& f''(y) - \beta_0(\gamma+1) y^\gamma f(y) = \lambda f(y) & &\mathrm{in }\, [0,1]\\
& f'(y) = 0 & &\mathrm{on }\, y=0, y=1,
\end{aligned}
\end{equation}
This is solved with high-resolution spectral methods with the Chebfun \cite{Driscoll2014} package. The eigenvalue $\lambda$ represents the effective homogeneous reaction that counterbalance the heterogeneous shear-induced events. This approach is also applicable to more general 2D-3D cases, giving a more stable and robust approach for deriving macroscopic reaction rates and (via homogenisation) transport equations.

\subsection{The FCC packing}
\begin{figure}[htbp]
\centering
\includegraphics[width=9cm, height=6.5cm]{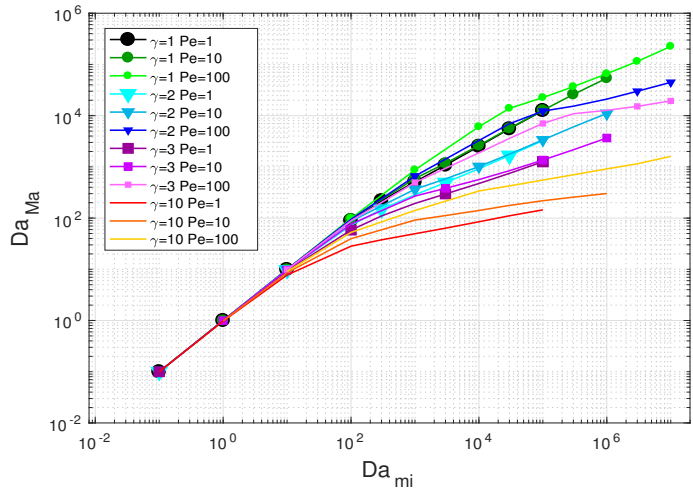}
\caption{FCC packing: macro- vs. micro-scale \Dam numbers.}
\label{fig:fcc}
\end{figure}

For a more complex three-dimensional FCC packing, as long as the process is diffusion controlled or the breakage frequency  is relatively small, the linear relation between the micro-scale \Dam and the macro-scale rate is recovered (similar to one shown in \cref{fig:channel}), as shown in \cref{fig:fcc}.

However, increasing the pore-scale breakage frequency, the results for this more realistic case are more complex that the ones presents for the CH system.
In particular, we can identify the following new features which were not present in \cref{fig:channel}: $i$) the transition from linear to power-law dependence spans a broader range of $\mathrm{Da}_{\mathrm{mi}}$ values with respect the two-dimensional channel, and it becomes more difficult to identify an asymptotic trend for each $\gamma$; $ii$) for $\mathrm{Da}_{\mathrm{mi}}>>1 $ a dependence on the \peclet number appears, and we can qualitatively infer from \cref{fig:fcc}  that this dependence increases with $\gamma$. 

Moreover, we can observe that transition regimes seem to extend up to values of $\da_{\mathrm{mi}}$ (i.e., $>10^{3}$) which are no longer realistic  for practical applications.
This can be explained in terms of influence of the porous medium structure on the reaction/transport phenomena. The dependence on the \peclet number seems to be larger for large values of $\gamma$ with a behaviour similar to the channel presented in the previous case when $\gamma=1$. The values of $\Da{mi}$ for which we have the switch from the linear to the non-linear behaviour has a dependence on $\gamma$ and $\pe$ and the switch happen at smaller $\Da{mi}$ number as the $\gamma$ and $\pe$ increase. 
This behaviour can at least intuitively explained in terms of influence of the porous medium structure on the reaction/transport phenomena. The higher is $\gamma$ the more heterogeneous is the reacting front, and the greater is the disturbance produced by the presence of the grains (while a sharper front would channelling through the medium more easily). 

The asymptotic behaviour with respect to $\gamma$, discovered for the channel, seems to persist also in this geometry, in particular for small $\pe$, while no definitive conclusions can be drawn for larger $\pe$ which could also be more sensitive to numerical errors.  Regardless of the geometry of the medium, the limiting case of $\gamma\to\infty$ is expected to behave like non-premixed combustion (in the mixing limited limited) and perfect deposition, where the reaction happens on a two-dimensional manifold only, and $\da_{\mathrm{Ma}}$ reaches a $\pe$-dependent flat value.

%
%

\section{Conclusions}

In this work we presented a model for particulate flows based on the Population Balance Equation (PBE) framework with specific consideration for micro-fluidics and porous media problems. We first introduced a novel dimensional analysis of the PBE including a new term describing  breakup after particle-particle collisions. We then focus on upscaling and the definition of averaged PBE kernels in heterogeneous flows.
Describing the particle dynamics and evolution in micro-scale complex geometries, in fact,  has enormous potential in a wide range of industrial and environmental applications, but also presents challenges which make such systems extremely interesting and at the same time extremely difficult to model. These include the prohibitive computational cost of fully resolved simulation and the need of macro/Darcy/continuum-scale models. The averaging operation, however, introduces additional coupling between the different particulate processes and between particles of different sizes. After describing a general upscaling framework  we limit the discussion to first-order shear-induced events, in particular the breakage of particles inside the pores. 
We showed how the shear-induced breakage can be considered as a linear process in the concentration of the particles and we derived the relevant dimensionless quantities to describe it at the pore- and macro-scales.
To  test our framework in numerical experiments, we considered a two dimensional channel, where simplified one-dimensional results can be obtained,  and a three-dimensional periodic porous medium made of a Face-Centred packing of spheres.

We showed that the upscaling procedure follows the expected behaviour, with the macroscopic breakage frequencies loosing the linear dependence with respect to its pore-scale analogue, due to the limited mixing. We also show the effect of the heterogeneity which is controlled by the exponent of the shear-induced breakage kernel. For the two-dimensional channel we propose a law for macroscopic effective reaction rate based on the pore-scale rate and the shear exponent. This qualitatively is reproduced also for more complicated system, like the sphere packing, but with a more complex dependence of the P\'eclet number.
This work represents a first attempt towards the systematic upscaling of the PBE framework in periodic porous media. Future work currently under development include the upscaling of non-linear processes as well as the interaction between particulate processes and deposition.

\section*{Acknowledgements}
This research has been funded by the Royal Academy of Engineering (Asphaltene dynamics at the pore-scale and the impact on oil production at the field-scale IAPP 18-19 285)

\bibliographystyle{apsrev}
\bibliography{bibliography_20Feb20}

\begin{thebibliography}{46}
\expandafter\ifx\csname natexlab\endcsname\relax\def\natexlab#1{#1}\fi
\expandafter\ifx\csname bibnamefont\endcsname\relax
  \def\bibnamefont#1{#1}\fi
\expandafter\ifx\csname bibfnamefont\endcsname\relax
  \def\bibfnamefont#1{#1}\fi
\expandafter\ifx\csname citenamefont\endcsname\relax
  \def\citenamefont#1{#1}\fi
\expandafter\ifx\csname url\endcsname\relax
  \def\url#1{\texttt{#1}}\fi
\expandafter\ifx\csname urlprefix\endcsname\relax\def\urlprefix{URL }\fi
\providecommand{\bibinfo}[2]{#2}
\providecommand{\eprint}[2][]{\url{#2}}

\bibitem[{\citenamefont{Ramkrishna}(2000)}]{Ramkrishna2000}
\bibinfo{author}{\bibfnamefont{D.}~\bibnamefont{Ramkrishna}},
  \emph{\bibinfo{title}{{Population Balances}}} (\bibinfo{publisher}{Academic
  Press: San Diego}, \bibinfo{year}{2000}).

\bibitem[{\citenamefont{Di~Pasquale et~al.}(2012)\citenamefont{Di~Pasquale,
  Marchisio, and Barresi}}]{DiPasquale2012}
\bibinfo{author}{\bibfnamefont{N.}~\bibnamefont{Di~Pasquale}},
  \bibinfo{author}{\bibfnamefont{D.}~\bibnamefont{Marchisio}},
  \bibnamefont{and} \bibinfo{author}{\bibfnamefont{A.}~\bibnamefont{Barresi}},
  \bibinfo{journal}{Chem. Eng. Sci.} \textbf{\bibinfo{volume}{84}},
  \bibinfo{pages}{671} (\bibinfo{year}{2012}).

\bibitem[{\citenamefont{Lavino et~al.}(2017)\citenamefont{Lavino, Di~Pasquale,
  Carbone, and Marchisio}}]{Lavino2017}
\bibinfo{author}{\bibfnamefont{A.~D.} \bibnamefont{Lavino}},
  \bibinfo{author}{\bibfnamefont{N.}~\bibnamefont{Di~Pasquale}},
  \bibinfo{author}{\bibfnamefont{P.}~\bibnamefont{Carbone}}, \bibnamefont{and}
  \bibinfo{author}{\bibfnamefont{D.~L.} \bibnamefont{Marchisio}},
  \bibinfo{journal}{Chemical Engineering Science}
  \textbf{\bibinfo{volume}{171}}, \bibinfo{pages}{485} (\bibinfo{year}{2017}).

\bibitem[{\citenamefont{Lattuada et~al.}(2016)\citenamefont{Lattuada, Zaccone,
  Wu, and Morbidelli}}]{Lattuada2016}
\bibinfo{author}{\bibfnamefont{M.}~\bibnamefont{Lattuada}},
  \bibinfo{author}{\bibfnamefont{A.}~\bibnamefont{Zaccone}},
  \bibinfo{author}{\bibfnamefont{H.}~\bibnamefont{Wu}}, \bibnamefont{and}
  \bibinfo{author}{\bibfnamefont{M.}~\bibnamefont{Morbidelli}},
  \bibinfo{journal}{Soft matter} \textbf{\bibinfo{volume}{12}},
  \bibinfo{pages}{5313} (\bibinfo{year}{2016}).

\bibitem[{\citenamefont{Sadegh-Vaziri et~al.}(2018)\citenamefont{Sadegh-Vaziri,
  Ludwig, Sundmacher, and Babler}}]{sadegh2018mechanisms}
\bibinfo{author}{\bibfnamefont{R.}~\bibnamefont{Sadegh-Vaziri}},
  \bibinfo{author}{\bibfnamefont{K.}~\bibnamefont{Ludwig}},
  \bibinfo{author}{\bibfnamefont{K.}~\bibnamefont{Sundmacher}},
  \bibnamefont{and} \bibinfo{author}{\bibfnamefont{M.~U.}
  \bibnamefont{Babler}}, \bibinfo{journal}{Journal of colloid and interface
  science} \textbf{\bibinfo{volume}{528}}, \bibinfo{pages}{336}
  (\bibinfo{year}{2018}).

\bibitem[{\citenamefont{Serra et~al.}(1997)\citenamefont{Serra, Colomer, and
  Casamitjana}}]{Serra1997}
\bibinfo{author}{\bibfnamefont{T.}~\bibnamefont{Serra}},
  \bibinfo{author}{\bibfnamefont{J.}~\bibnamefont{Colomer}}, \bibnamefont{and}
  \bibinfo{author}{\bibfnamefont{X.}~\bibnamefont{Casamitjana}},
  \bibinfo{journal}{Journal of Colloid and Interface Science}
  \textbf{\bibinfo{volume}{187}}, \bibinfo{pages}{466} (\bibinfo{year}{1997}).

\bibitem[{\citenamefont{Serra and Casamitjana}(1998)}]{Serra1998}
\bibinfo{author}{\bibfnamefont{T.}~\bibnamefont{Serra}} \bibnamefont{and}
  \bibinfo{author}{\bibfnamefont{X.}~\bibnamefont{Casamitjana}},
  \bibinfo{journal}{AIChE journal} \textbf{\bibinfo{volume}{44}},
  \bibinfo{pages}{1724} (\bibinfo{year}{1998}).

\bibitem[{\citenamefont{Zitha and Du}(2010)}]{Zitha2010}
\bibinfo{author}{\bibfnamefont{P.~L.~J.} \bibnamefont{Zitha}} \bibnamefont{and}
  \bibinfo{author}{\bibfnamefont{D.~X.} \bibnamefont{Du}},
  \bibinfo{journal}{Transport in Porous Media} \textbf{\bibinfo{volume}{83}},
  \bibinfo{pages}{603} (\bibinfo{year}{2010}).

\bibitem[{\citenamefont{Buffo et~al.}(2012)\citenamefont{Buffo, Vanni, and
  Marchisio}}]{Buffo2012}
\bibinfo{author}{\bibfnamefont{A.}~\bibnamefont{Buffo}},
  \bibinfo{author}{\bibfnamefont{M.}~\bibnamefont{Vanni}}, \bibnamefont{and}
  \bibinfo{author}{\bibfnamefont{D.}~\bibnamefont{Marchisio}},
  \bibinfo{journal}{Chemical Engineering Science}
  \textbf{\bibinfo{volume}{70}}, \bibinfo{pages}{31} (\bibinfo{year}{2012}).

\bibitem[{\citenamefont{Di~Pasquale et~al.}(2013)\citenamefont{Di~Pasquale,
  Marchisio, Carbone, and Barresi}}]{DiPasquale2013}
\bibinfo{author}{\bibfnamefont{N.}~\bibnamefont{Di~Pasquale}},
  \bibinfo{author}{\bibfnamefont{D.~L.} \bibnamefont{Marchisio}},
  \bibinfo{author}{\bibfnamefont{P.}~\bibnamefont{Carbone}}, \bibnamefont{and}
  \bibinfo{author}{\bibfnamefont{A.~A.} \bibnamefont{Barresi}},
  \bibinfo{journal}{Chem. Eng. Res. Des.} \textbf{\bibinfo{volume}{91}},
  \bibinfo{pages}{2275} (\bibinfo{year}{2013}).

\bibitem[{\citenamefont{Di~Pasquale et~al.}(2014)\citenamefont{Di~Pasquale,
  Marchisio, Barresi, and Carbone}}]{DiPasquale2014pcl}
\bibinfo{author}{\bibfnamefont{N.}~\bibnamefont{Di~Pasquale}},
  \bibinfo{author}{\bibfnamefont{D.~L.} \bibnamefont{Marchisio}},
  \bibinfo{author}{\bibfnamefont{A.~A.} \bibnamefont{Barresi}},
  \bibnamefont{and} \bibinfo{author}{\bibfnamefont{P.}~\bibnamefont{Carbone}},
  \bibinfo{journal}{Journal of Physical Chemistry B}
  \textbf{\bibinfo{volume}{118}}, \bibinfo{pages}{13258}
  (\bibinfo{year}{2014}).

\bibitem[{\citenamefont{Frungieri et~al.}(2020)\citenamefont{Frungieri, Babler,
  and Vanni}}]{frungieri2020shear}
\bibinfo{author}{\bibfnamefont{G.}~\bibnamefont{Frungieri}},
  \bibinfo{author}{\bibfnamefont{M.~U.} \bibnamefont{Babler}},
  \bibnamefont{and} \bibinfo{author}{\bibfnamefont{M.}~\bibnamefont{Vanni}},
  \bibinfo{journal}{Langmuir} \textbf{\bibinfo{volume}{36}},
  \bibinfo{pages}{10739} (\bibinfo{year}{2020}).

\bibitem[{\citenamefont{Marcato et~al.}(2021)\citenamefont{Marcato, Boccardo,
  and Marchisio}}]{Marcato2021}
\bibinfo{author}{\bibfnamefont{A.}~\bibnamefont{Marcato}},
  \bibinfo{author}{\bibfnamefont{G.}~\bibnamefont{Boccardo}}, \bibnamefont{and}
  \bibinfo{author}{\bibfnamefont{D.}~\bibnamefont{Marchisio}},
  \bibinfo{journal}{Chemical Engineering Journal}
  \textbf{\bibinfo{volume}{417}}, \bibinfo{pages}{128936}
  (\bibinfo{year}{2021}), ISSN \bibinfo{issn}{1385-8947},
  \urlprefix\url{https://www.sciencedirect.com/science/article/pii/S1385894721005295}.

\bibitem[{\citenamefont{Boccardo et~al.}(2017)\citenamefont{Boccardo,
  Crevacore, Sethi, and Icardi}}]{Boccardo2017}
\bibinfo{author}{\bibfnamefont{G.}~\bibnamefont{Boccardo}},
  \bibinfo{author}{\bibfnamefont{E.}~\bibnamefont{Crevacore}},
  \bibinfo{author}{\bibfnamefont{R.}~\bibnamefont{Sethi}}, \bibnamefont{and}
  \bibinfo{author}{\bibfnamefont{M.}~\bibnamefont{Icardi}},
  \bibinfo{journal}{Journal of contaminant hydrology}
  \textbf{\bibinfo{volume}{212}}, \bibinfo{pages}{3} (\bibinfo{year}{2017}).

\bibitem[{\citenamefont{Icardi and Dentz}(2020)}]{Icardi2020}
\bibinfo{author}{\bibfnamefont{M.}~\bibnamefont{Icardi}} \bibnamefont{and}
  \bibinfo{author}{\bibfnamefont{M.}~\bibnamefont{Dentz}},
  \bibinfo{journal}{GEM-International Journal on Geomathematics}
  \textbf{\bibinfo{volume}{11}}, \bibinfo{pages}{1} (\bibinfo{year}{2020}).

\bibitem[{\citenamefont{Krehel et~al.}(2015)\citenamefont{Krehel, Muntean, and
  Knabner}}]{Krehel2015}
\bibinfo{author}{\bibfnamefont{O.}~\bibnamefont{Krehel}},
  \bibinfo{author}{\bibfnamefont{A.}~\bibnamefont{Muntean}}, \bibnamefont{and}
  \bibinfo{author}{\bibfnamefont{P.}~\bibnamefont{Knabner}},
  \bibinfo{journal}{Adv. Water Resour.} \textbf{\bibinfo{volume}{86}},
  \bibinfo{pages}{209} (\bibinfo{year}{2015}), ISSN \bibinfo{issn}{03091708},
  \eprint{1404.4207},
  \urlprefix\url{http://dx.doi.org/10.1016/j.advwatres.2015.10.005}.

\bibitem[{\citenamefont{Won et~al.}(2021)\citenamefont{Won, Lee, and
  Burns}}]{won2021upscaling}
\bibinfo{author}{\bibfnamefont{J.}~\bibnamefont{Won}},
  \bibinfo{author}{\bibfnamefont{J.}~\bibnamefont{Lee}}, \bibnamefont{and}
  \bibinfo{author}{\bibfnamefont{S.~E.} \bibnamefont{Burns}},
  \bibinfo{journal}{Acta Geotechnica} \textbf{\bibinfo{volume}{16}},
  \bibinfo{pages}{421} (\bibinfo{year}{2021}).

\bibitem[{\citenamefont{Boccardo et~al.}(2019)\citenamefont{Boccardo, Sethi,
  and Marchisio}}]{Boccardo2019}
\bibinfo{author}{\bibfnamefont{G.}~\bibnamefont{Boccardo}},
  \bibinfo{author}{\bibfnamefont{R.}~\bibnamefont{Sethi}}, \bibnamefont{and}
  \bibinfo{author}{\bibfnamefont{D.~L.} \bibnamefont{Marchisio}},
  \bibinfo{journal}{Chemical Engineering Science}
  \textbf{\bibinfo{volume}{198}}, \bibinfo{pages}{290} (\bibinfo{year}{2019}).

\bibitem[{\citenamefont{Seetha et~al.}(2017)\citenamefont{Seetha, Raoof, Kumar,
  and Hassanizadeh}}]{seetha2017upscaling}
\bibinfo{author}{\bibfnamefont{N.}~\bibnamefont{Seetha}},
  \bibinfo{author}{\bibfnamefont{A.}~\bibnamefont{Raoof}},
  \bibinfo{author}{\bibfnamefont{M.~M.} \bibnamefont{Kumar}}, \bibnamefont{and}
  \bibinfo{author}{\bibfnamefont{S.~M.} \bibnamefont{Hassanizadeh}},
  \bibinfo{journal}{Journal of contaminant hydrology}
  \textbf{\bibinfo{volume}{200}}, \bibinfo{pages}{1} (\bibinfo{year}{2017}).

\bibitem[{\citenamefont{Bedrikovetsky}(2008)}]{bedrikovetsky2008upscaling}
\bibinfo{author}{\bibfnamefont{P.}~\bibnamefont{Bedrikovetsky}},
  \bibinfo{journal}{Transport in Porous Media} \textbf{\bibinfo{volume}{75}},
  \bibinfo{pages}{335} (\bibinfo{year}{2008}).

\bibitem[{\citenamefont{Nassar et~al.}(2015)\citenamefont{Nassar, Betancur,
  Acevedo, Franco, and Cort{\'e}s}}]{nassar2015development}
\bibinfo{author}{\bibfnamefont{N.~N.} \bibnamefont{Nassar}},
  \bibinfo{author}{\bibfnamefont{S.}~\bibnamefont{Betancur}},
  \bibinfo{author}{\bibfnamefont{S.}~\bibnamefont{Acevedo}},
  \bibinfo{author}{\bibfnamefont{C.~A.} \bibnamefont{Franco}},
  \bibnamefont{and} \bibinfo{author}{\bibfnamefont{F.~B.}
  \bibnamefont{Cort{\'e}s}}, \bibinfo{journal}{Industrial \& Engineering
  Chemistry Research} \textbf{\bibinfo{volume}{54}}, \bibinfo{pages}{8201}
  (\bibinfo{year}{2015}).

\bibitem[{\citenamefont{Patzek}(1988)}]{Patzek1988}
\bibinfo{author}{\bibfnamefont{T.~W.} \bibnamefont{Patzek}},
  \emph{\bibinfo{title}{Description of Foam Flow in Porous Media by the
  Population Balance Method}} (\bibinfo{year}{1988}),
  chap.~\bibinfo{chapter}{16}, pp. \bibinfo{pages}{326--341},
  \urlprefix\url{https://pubs.acs.org/doi/abs/10.1021/bk-1988-0373.ch016}.

\bibitem[{\citenamefont{Auriault and Adler}(1995)}]{auriault1995taylor}
\bibinfo{author}{\bibfnamefont{J.~L.} \bibnamefont{Auriault}} \bibnamefont{and}
  \bibinfo{author}{\bibfnamefont{P.~M.} \bibnamefont{Adler}},
  \bibinfo{journal}{Advances in Water Resources} \textbf{\bibinfo{volume}{18}},
  \bibinfo{pages}{217} (\bibinfo{year}{1995}).

\bibitem[{\citenamefont{Marchisio and Fox}(2013)}]{Marchisio2013}
\bibinfo{author}{\bibfnamefont{D.~L.} \bibnamefont{Marchisio}}
  \bibnamefont{and} \bibinfo{author}{\bibfnamefont{R.~O.} \bibnamefont{Fox}},
  \emph{\bibinfo{title}{Computational models for polydisperse particulate and
  multiphase systems}} (\bibinfo{publisher}{Cambridge University Press},
  \bibinfo{year}{2013}).

\bibitem[{\citenamefont{Ferry et~al.}(2003)\citenamefont{Ferry, Rani, and
  Balachandar}}]{Ferry2003}
\bibinfo{author}{\bibfnamefont{J.}~\bibnamefont{Ferry}},
  \bibinfo{author}{\bibfnamefont{S.~L.} \bibnamefont{Rani}}, \bibnamefont{and}
  \bibinfo{author}{\bibfnamefont{S.}~\bibnamefont{Balachandar}},
  \bibinfo{journal}{International journal of multiphase flow}
  \textbf{\bibinfo{volume}{29}}, \bibinfo{pages}{869} (\bibinfo{year}{2003}).

\bibitem[{\citenamefont{Bird et~al.}(2002)\citenamefont{Bird, Stewart, and
  Lightfoot}}]{Bird2002}
\bibinfo{author}{\bibfnamefont{R.}~\bibnamefont{Bird}},
  \bibinfo{author}{\bibfnamefont{W.}~\bibnamefont{Stewart}}, \bibnamefont{and}
  \bibinfo{author}{\bibfnamefont{E.}~\bibnamefont{Lightfoot}},
  \emph{\bibinfo{title}{{Transport Phenomena}}} (\bibinfo{publisher}{{John
  Wiley} \& {Sons, Inc.}: New York}, \bibinfo{year}{2002}).

\bibitem[{\citenamefont{Chen and Li}(2020)}]{chen2020collision}
\bibinfo{author}{\bibfnamefont{S.}~\bibnamefont{Chen}} \bibnamefont{and}
  \bibinfo{author}{\bibfnamefont{S.}~\bibnamefont{Li}},
  \bibinfo{journal}{Journal of Fluid Mechanics} \textbf{\bibinfo{volume}{902}}
  (\bibinfo{year}{2020}).

\bibitem[{\citenamefont{Khalifa and Breuer}(2021)}]{khalifa2021efficient}
\bibinfo{author}{\bibfnamefont{A.}~\bibnamefont{Khalifa}} \bibnamefont{and}
  \bibinfo{author}{\bibfnamefont{M.}~\bibnamefont{Breuer}},
  \bibinfo{journal}{International Journal of Multiphase Flow}
  \textbf{\bibinfo{volume}{142}}, \bibinfo{pages}{103625}
  (\bibinfo{year}{2021}).

\bibitem[{\citenamefont{Frungieri and Vanni}(2021)}]{Frungieri2021}
\bibinfo{author}{\bibfnamefont{G.}~\bibnamefont{Frungieri}} \bibnamefont{and}
  \bibinfo{author}{\bibfnamefont{M.}~\bibnamefont{Vanni}},
  \bibinfo{journal}{Powder Technology} \textbf{\bibinfo{volume}{388}},
  \bibinfo{pages}{357} (\bibinfo{year}{2021}).

\bibitem[{\citenamefont{Barthelmes et~al.}(2003)\citenamefont{Barthelmes,
  Pratsinis, and Buggisch}}]{Barthelmes2003}
\bibinfo{author}{\bibfnamefont{G.}~\bibnamefont{Barthelmes}},
  \bibinfo{author}{\bibfnamefont{S.~E.} \bibnamefont{Pratsinis}},
  \bibnamefont{and} \bibinfo{author}{\bibfnamefont{H.}~\bibnamefont{Buggisch}},
  \bibinfo{journal}{Chemical Engineering Science}
  \textbf{\bibinfo{volume}{58}}, \bibinfo{pages}{2893} (\bibinfo{year}{2003}).

\bibitem[{\citenamefont{Solsvik et~al.}(2013)\citenamefont{Solsvik, Tangen, and
  Jakobsen}}]{Solsvik2013}
\bibinfo{author}{\bibfnamefont{J.}~\bibnamefont{Solsvik}},
  \bibinfo{author}{\bibfnamefont{S.}~\bibnamefont{Tangen}}, \bibnamefont{and}
  \bibinfo{author}{\bibfnamefont{H.~A.} \bibnamefont{Jakobsen}},
  \bibinfo{journal}{Reviews in Chemical Engineering}
  \textbf{\bibinfo{volume}{29}}, \bibinfo{pages}{241} (\bibinfo{year}{2013}).

\bibitem[{\citenamefont{Vanni}(2000)}]{Vanni2000}
\bibinfo{author}{\bibfnamefont{M.}~\bibnamefont{Vanni}},
  \bibinfo{journal}{Journal of colloid and interface science}
  \textbf{\bibinfo{volume}{221}}, \bibinfo{pages}{143} (\bibinfo{year}{2000}).

\bibitem[{\citenamefont{Grigoriev et~al.}(2020)\citenamefont{Grigoriev, Iliev,
  and Vabishchevich}}]{Grigoriev2020}
\bibinfo{author}{\bibfnamefont{V.~V.} \bibnamefont{Grigoriev}},
  \bibinfo{author}{\bibfnamefont{O.}~\bibnamefont{Iliev}}, \bibnamefont{and}
  \bibinfo{author}{\bibfnamefont{P.~N.} \bibnamefont{Vabishchevich}},
  \bibinfo{journal}{Journal of Computational and Applied Mathematics}
  \textbf{\bibinfo{volume}{370}}, \bibinfo{pages}{112661}
  (\bibinfo{year}{2020}), ISSN \bibinfo{issn}{0377-0427},
  \urlprefix\url{https://www.sciencedirect.com/science/article/pii/S0377042719306661}.

\bibitem[{\citenamefont{Municchi et~al.}(-)\citenamefont{Municchi, Boccardo,
  Tartakovsky, and Icardi}}]{fastreaction}
\bibinfo{author}{\bibfnamefont{F.}~\bibnamefont{Municchi}},
  \bibinfo{author}{\bibfnamefont{G.}~\bibnamefont{Boccardo}},
  \bibinfo{author}{\bibfnamefont{D.}~\bibnamefont{Tartakovsky}},
  \bibnamefont{and} \bibinfo{author}{\bibfnamefont{M.}~\bibnamefont{Icardi}},
  \bibinfo{journal}{in preparation}  (\bibinfo{year}{-}).

\bibitem[{\citenamefont{Whitaker}(1998)}]{whitaker1998}
\bibinfo{author}{\bibfnamefont{S.}~\bibnamefont{Whitaker}},
  \emph{\bibinfo{title}{The method of volume averaging}},
  vol.~\bibinfo{volume}{13} (\bibinfo{publisher}{Springer Science \& Business
  Media}, \bibinfo{year}{1998}).

\bibitem[{\citenamefont{Hornung}(1991)}]{Hornung}
\bibinfo{author}{\bibfnamefont{U.}~\bibnamefont{Hornung}},
  \emph{\bibinfo{title}{{Homogenization and Porous Media}}}
  (\bibinfo{year}{1991}), ISBN \bibinfo{isbn}{0-387-94786-8}.

\bibitem[{\citenamefont{Saha et~al.}(2016)\citenamefont{Saha, Babler, Holzner,
  Soos, L{\"u}thi, Liberzon, and Kinzelbach}}]{saha2016breakup}
\bibinfo{author}{\bibfnamefont{D.}~\bibnamefont{Saha}},
  \bibinfo{author}{\bibfnamefont{M.~U.} \bibnamefont{Babler}},
  \bibinfo{author}{\bibfnamefont{M.}~\bibnamefont{Holzner}},
  \bibinfo{author}{\bibfnamefont{M.}~\bibnamefont{Soos}},
  \bibinfo{author}{\bibfnamefont{B.}~\bibnamefont{L{\"u}thi}},
  \bibinfo{author}{\bibfnamefont{A.}~\bibnamefont{Liberzon}}, \bibnamefont{and}
  \bibinfo{author}{\bibfnamefont{W.}~\bibnamefont{Kinzelbach}},
  \bibinfo{journal}{Langmuir} \textbf{\bibinfo{volume}{32}},
  \bibinfo{pages}{55} (\bibinfo{year}{2016}).

\bibitem[{\citenamefont{Elimelech et~al.}(1995)\citenamefont{Elimelech,
  Gregory, Jia, and Williams}}]{Elimelech1995}
\bibinfo{author}{\bibfnamefont{M.}~\bibnamefont{Elimelech}},
  \bibinfo{author}{\bibfnamefont{J.}~\bibnamefont{Gregory}},
  \bibinfo{author}{\bibfnamefont{X.}~\bibnamefont{Jia}}, \bibnamefont{and}
  \bibinfo{author}{\bibfnamefont{R.}~\bibnamefont{Williams}},
  \emph{\bibinfo{title}{{Particle Deposition \& Aggregation}}}
  (\bibinfo{publisher}{{Butterwort-Heinemann}}, \bibinfo{year}{1995}).

\bibitem[{\citenamefont{Boccardo et~al.}(2018)\citenamefont{Boccardo, Sokolov,
  and Paster}}]{Boccardo2018}
\bibinfo{author}{\bibfnamefont{G.}~\bibnamefont{Boccardo}},
  \bibinfo{author}{\bibfnamefont{I.~M.} \bibnamefont{Sokolov}},
  \bibnamefont{and} \bibinfo{author}{\bibfnamefont{A.}~\bibnamefont{Paster}},
  \bibinfo{journal}{Journal of Computational Physics}
  \textbf{\bibinfo{volume}{374}}, \bibinfo{pages}{1152} (\bibinfo{year}{2018}).

\bibitem[{\citenamefont{Allaire and Raphael}(2007)}]{Allaire2007}
\bibinfo{author}{\bibfnamefont{G.}~\bibnamefont{Allaire}} \bibnamefont{and}
  \bibinfo{author}{\bibfnamefont{A.-L.} \bibnamefont{Raphael}},
  \bibinfo{journal}{Comptes Rendus Mathematique}
  \textbf{\bibinfo{volume}{344}}, \bibinfo{pages}{523} (\bibinfo{year}{2007}).

\bibitem[{\citenamefont{Mauri}(1991)}]{Mauri1991}
\bibinfo{author}{\bibfnamefont{R.}~\bibnamefont{Mauri}},
  \bibinfo{journal}{Physics of Fluids A: Fluid Dynamics}
  \textbf{\bibinfo{volume}{3}}, \bibinfo{pages}{743} (\bibinfo{year}{1991}).

\bibitem[{\citenamefont{Municchi and Icardi}(2020)}]{Municchi2020}
\bibinfo{author}{\bibfnamefont{F.}~\bibnamefont{Municchi}} \bibnamefont{and}
  \bibinfo{author}{\bibfnamefont{M.}~\bibnamefont{Icardi}},
  \bibinfo{journal}{Physical Review Research} \textbf{\bibinfo{volume}{2}},
  \bibinfo{pages}{013041} (\bibinfo{year}{2020}).

\bibitem[{\citenamefont{Crevacore et~al.}(2016)\citenamefont{Crevacore,
  Boccardo, Marchisio, and Sethi}}]{Crevacore2016}
\bibinfo{author}{\bibfnamefont{E.}~\bibnamefont{Crevacore}},
  \bibinfo{author}{\bibfnamefont{G.}~\bibnamefont{Boccardo}},
  \bibinfo{author}{\bibfnamefont{D.}~\bibnamefont{Marchisio}},
  \bibnamefont{and} \bibinfo{author}{\bibfnamefont{R.}~\bibnamefont{Sethi}},
  \bibinfo{journal}{Chemical Engineering Transactions}
  \textbf{\bibinfo{volume}{47}}, \bibinfo{pages}{271} (\bibinfo{year}{2016}).

\bibitem[{\citenamefont{Boccardo et~al.}(2020)\citenamefont{Boccardo,
  Crevacore, Passalacqua, and Icardi}}]{Boccardo2020}
\bibinfo{author}{\bibfnamefont{G.}~\bibnamefont{Boccardo}},
  \bibinfo{author}{\bibfnamefont{E.}~\bibnamefont{Crevacore}},
  \bibinfo{author}{\bibfnamefont{A.}~\bibnamefont{Passalacqua}},
  \bibnamefont{and} \bibinfo{author}{\bibfnamefont{M.}~\bibnamefont{Icardi}},
  \bibinfo{journal}{Computing and Visualization in Science}
  \textbf{\bibinfo{volume}{23}} (\bibinfo{year}{2020}).

\bibitem[{\citenamefont{Dentz et~al.}(2018)\citenamefont{Dentz, Icardi, and
  Hidalgo}}]{dentz2018mechanisms}
\bibinfo{author}{\bibfnamefont{M.}~\bibnamefont{Dentz}},
  \bibinfo{author}{\bibfnamefont{M.}~\bibnamefont{Icardi}}, \bibnamefont{and}
  \bibinfo{author}{\bibfnamefont{J.~J.} \bibnamefont{Hidalgo}},
  \bibinfo{journal}{Journal of Fluid Mechanics} \textbf{\bibinfo{volume}{841}},
  \bibinfo{pages}{851} (\bibinfo{year}{2018}).

\bibitem[{\citenamefont{Driscoll et~al.}(2014)\citenamefont{Driscoll, Hale, and
  Trefethen}}]{Driscoll2014}
\bibinfo{author}{\bibfnamefont{T.~A.} \bibnamefont{Driscoll}},
  \bibinfo{author}{\bibfnamefont{N.}~\bibnamefont{Hale}}, \bibnamefont{and}
  \bibinfo{author}{\bibfnamefont{L.~N.} \bibnamefont{Trefethen}},
  \emph{\bibinfo{title}{Chebfun Guide}} (\bibinfo{publisher}{Pafnuty
  Publications}, \bibinfo{year}{2014}),
  \urlprefix\url{http://www.chebfun.org/docs/guide/}.

\end{thebibliography}

\end{document}